\documentclass[reprint,amsmath,amssymb,aps,superscriptaddress,floatfix,prmaterials,longbibliography]{revtex4-2}

\usepackage{graphicx}
\usepackage{dcolumn}
\usepackage{bm}
\usepackage{color}
\usepackage[caption=false]{subfig}
\usepackage{booktabs}
\usepackage{gensymb}

\newcommand*\UCG{UCr$_6$Ge$_6$}
\newcommand*\LCG{LuCr$_6$Ge$_6$}

\begin{document}

\title{Structural modulation, physical properties, and electronic band structure of the kagome metal \UCG}

\author{Z.~W.~Riedel}
\affiliation{Los Alamos National Laboratory, Los Alamos, New Mexico 87545, USA}

\author{{\color{black}P.~A.~E.~Murgatroyd}}
\affiliation{Los Alamos National Laboratory, Los Alamos, New Mexico 87545, USA}

\author{C.~S.~Kengle}
\affiliation{Los Alamos National Laboratory, Los Alamos, New Mexico 87545, USA}

\author{{\color{black}P.~M.~T.~Vianez}}
\affiliation{Los Alamos National Laboratory, Los Alamos, New Mexico 87545, USA}

\author{A.~Schmidt}
\affiliation{Bruker AXS, Madison, Wisconsin 53711, USA}

\author{{\color{black}X.~Du}}
\affiliation{{\color{black}Department of Applied Physics, Yale University, New Haven, Connecticut 06511, USA}}

\author{K.~Allen}
\affiliation{Los Alamos National Laboratory, Los Alamos, New Mexico 87545, USA}
\affiliation{
Department of Physics and
Astronomy, Rice University, Houston, Texas 77005, USA}

\author{{\color{black}T.~K.~Kim}}
\affiliation{{\color{black}Diamond Light Source Ltd., Harwell Science and Innovation Campus, Didcot, OX110DE, UK}}

\author{C.~Lane}
\affiliation{Los Alamos National Laboratory, Los Alamos, New Mexico 87545, USA}

\author{Ying Wai Li}
\affiliation{Los Alamos National Laboratory, Los Alamos, New Mexico 87545, USA}

\author{Jian-Xin Zhu}
\affiliation{Los Alamos National Laboratory, Los Alamos, New Mexico 87545, USA}

\author{J.~D.~Thompson}
\affiliation{Los Alamos National Laboratory, Los Alamos, New Mexico 87545, USA}

\author{F.~Ronning}
\affiliation{Los Alamos National Laboratory, Los Alamos, New Mexico 87545, USA}

\author{S.~M.~Thomas}
\affiliation{Los Alamos National Laboratory, Los Alamos, New Mexico 87545, USA}

\author{P.~F.~S.~Rosa}
\affiliation{Los Alamos National Laboratory, Los Alamos, New Mexico 87545, USA}

\author{E.~D.~Bauer}
\affiliation{Los Alamos National Laboratory, Los Alamos, New Mexico 87545, USA}

\begin{abstract}
The chemical flexibility of the $RM_6X_6$ stoichiometry, where an $f$-block element is intercalated in the CoSn structure type, allows for the tuning of flatbands associated with kagome lattices to the Fermi level and for emergent phenomena due to interactions between the $f$- and $d$-electron lattices. Yet, 5$f$ members of the ``166" compounds are underrepresented compared with 4$f$ members. Here, we report single-crystal growth of \UCG, which crystallizes in a monoclinically distorted Y$_{0.5}$Co$_3$Ge$_3$-type structure. The real-space character of the modulation, which is unique within the $RM_6X_6$ family, is approximated by a 3$\times$1$\times$2 supercell of the average monoclinic cell. The compound has kagome-lattice flatbands near the Fermi level and a moderately enhanced electronic heat capacity, as evidenced by its low-temperature Sommerfeld coefficient ($\gamma=86.5$~mJ~mol$^{-1}$~K$^{-2}$) paired with band structure calculations. The small, isotropic magnetization and featureless resistivity of \UCG\ suggest itinerant uranium 5$f$ electrons and Pauli paramagnetism. {\color{black}Angle-resolved photoemission spectroscopy results provide evidence for uranium 5$f$ weight at the Fermi level and for a flatband near the Fermi level associated with the chromium $3d$ kagome lattice.} The isotropic magnetic behavior of the uranium 5$f$ electrons starkly contrasts with localized behavior in other uranium 166 compounds, highlighting the high tunability of the magnetic ground state across the material family.
\end{abstract}

\maketitle

\section{Introduction}
Materials with the $RM_6X_6$ stoichiometry ($R$=Sc, Y, $f$-block element; $M$=transition metal; $X$=Ga, Si, Ge, Sn) contain a kagome transition metal lattice intercalated by an $R$ element. The interplay of these sublattices often produces emergent phenomena. For example, in $RM_6X_6$ (``166") materials with small $R$ elements relative to the $M-X$ cages surrounding them, charge density wave instabilities can occur, as in ScV$_6$Sn$_6$ and LuNb$_6$Sn$_6$ \cite{arachchige2022charge,tuniz2023dynamics,pokharel2023frustrated,kim2023infrared,meier2023tiny,hu2024phonon,yi2024tuning,ortiz2024stability,meier2025pressure}. In systems with magnetic $R$ or $M$ elements, anisotropic magnetic behavior may produce complex magnetic phase diagrams involving incommensurate phases or noncollinear spin structures that produce large topological Hall effects \cite{clatterbuck1999magnetic,kimura2006high,riberolles2024new,ghimire2020competing,l2025high,dhakal2021anisotropically,riedel2025magnetic}. Moreover, the extensive chemical tunability of the 166 family offers a platform for controlling electronic and structural behavior, such as Fermi level ($E_\mathrm{F}$) density of states (DOS) enhancement due to kagome-lattice flatbands or reduced dimensionality due to well-separated kagome layers. However, despite numerous reported 166 materials, most show no more than a moderate enhancement in the DOS at $E_\mathrm{F}$, and only a handful have reported Sommerfeld coefficients exceeding 60~mJ~mol$^{-1}$~K$^{-2}$ \cite{pokharel2021electronic,ishii2013ycr6ge6,avila2005direct,lyu2024anomalous,shi2025weak,pokharel2022highly,guo2023triangular,lou2025strongly}, all in the $R$V$_6$Sn$_6$, $R$Cr$_6$Ge$_6$, $R$Mn$_6$Sn$_6$, $R$Fe$_6$Ge$_6$, and $R$Fe$_6$Sn$_6$ subfamilies.

Reports of actinide 166 materials are sparse compared with those for lanthanide 166s. Therefore, actinides' 5$f$ electrons offer a new route to tune electronic behavior associated with kagome-lattice flatbands. 
Reported actinide 166s include UCo$_6$Ge$_6$ \cite{buchholz1981intermetallische}, UFe$_6$Ge$_6$ \cite{gonccalves1994ufe6ge6,waerenborgh2005crystal}, UV$_6$Sn$_6$ \cite{thomas2025uv6sn6,patino2025incom}, ThV$_6$Sn$_6$ \cite{thomas2025uv6sn6,xiao2024preparation}, UNb$_6$Sn$_6$ \cite{riedel2025magnetic}, and ThNb$_6$Sn$_6$ \cite{riedel2025magnetic}.
Recent work on UV$_6$Sn$_6$ and UNb$_6$Sn$_6$ has shown that the $d$-electron flatband is closer to $E_\mathrm{F}$ and less dispersive for UV$_6$Sn$_6$ \cite{riedel2025magnetic,thomas2025uv6sn6}, indicating that chemical substitutions may significantly tune the energy level and shape of the kagome-lattice flatband.
Here, we report a new uranium member of the $R$Cr$_6$Ge$_6$ subfamily \cite{mulder1993Gd,schobinger1997atomic,schobinger1997ferrimagnetism,konyk2020electrical,yang2024crystal,romaka2024structure,lee2025coexisting}, which includes YCr$_6$Ge$_6$ and YbCr$_6$Ge$_6$, two compounds with kagome-lattice flatbands close to $E_\mathrm{F}$ in the calculated band structures {\color{black}and ARPES data} \cite{ishii2013ycr6ge6,wang2020experimental,lee2025coexisting,lv2026cooperative}. We report on the single-crystal synthesis and physical properties of \UCG, a compound with unique magnetic behavior and the largest electronic heat capacity among the reported actinide 166s {\color{black}($\gamma=86.5$~mJ~mol$^{-1}$~K$^{-2}$)}.

\section{Methods}
\subsection{Experimental}
Single crystals of \UCG\ were grown from tin flux. Depleted uranium (99.99\%), chromium (Thermo Scientific, 99.997\%), germanium (Thermo Scientific, 99.9999+\%), and tin (Thermo Scientific, 99.999\%) pieces were placed in an alumina crucible in a 1:6:18:100 or 1.5:6:18:100 molar ratio. Samples were heated to 1100$\degree$C at 100$\degree$C/h, homogenized at 1100$\degree$C for 72~h, and then slow cooled at 2$\degree$C/h to 800$\degree$C, where they were centrifuged to remove the tin flux. For the 1.5:6:18:100 ratio, hexagonal plate crystals were recovered. For the 1:6:18:100 ratio, hexagonal plates as well as hexagonal rods were recovered. Additionally, needle-shaped single crystals of \LCG\ were grown in tin flux using a 1:6:6:20 (Lu:Cr:Ge:Sn) molar ratio and were spun at 500$^\circ$C, with LuCr$_x$Ge$_2$ crystals forming as a by-product (CeNiSi$_2$-type, $Cmcm$).

The crystal structure of \UCG\ was determined using data collected with a Bruker D8 VENTURE KAPPA single crystal X-ray diffractometer with an I$\mu$S 3.0 microfocus source ($\lambda=$ 0.71073~\AA), a HELIOS optics monocromator, and a PHOTON II CPAD detector. All data were integrated with \textsc{SAINT} V8.42 \cite{saint2025bruker}, yielding 5733 reflections, of which 311 where independent (average redundancy 18.43) and 100\% were greater than 2$\sigma$(F$^2$). A Multi-Scan absorption correction using \textsc{SADABS}-2016/2 \cite{krause2015comparison} was applied. The structure was solved by Intrinsic Phasing methods with \textsc{SHELXT}-2018/2 \cite{sheldrick2015shelxt} and refined by full-matrix least-squares methods against F$^2$ using \textsc{SHELXL}-2019/2 \cite{sheldrick2015crystal}. All atoms were refined with anisotropic displacement parameters. Single crystal diffraction of \LCG\ followed the same general procedure. Crystallographic data for the structures reported in this paper have been deposited with the Cambridge Crystallographic Data Centre \cite{groom2016cambridge}, and CIFs are provided in the Supplemental Material \cite{supplement}.

To collect heat capacity data, the two-tau thermal relaxation method was used with a Quantum Design Physical Property Measurement System (PPMS). A $^3$He fridge attachment was used for data collected below 1.8~K. For resistivity measurements below 400~K and up to 9~T, platinum wires in a four-point configuration were attached to \UCG\ crystals by spot welding. The crystal orientation was confirmed with a Photonic Science Laue diffractometer. The kagome plane was found to be parallel to the face of the plate-like crystals and perpendicular to the long axis of the rod-like crystals. Data collected under a magnetic field were symmetrized at positive and negative fields [($R_{H>0}$ + $R_{H<0}$)/2] to account for any small offset between the voltage leads. {\color{black}Hall resistance data were also collected on a plate-like crystal, and data were antisymmetrized using [($R_{H>0}$ - $R_{H<0}$)/2].}
Energy dispersive spectroscopy (EDS) maps were collected with a ThermoFisher Apreo 2 S scanning electron microscope and were processed with the Oxford program \textsc{AZtec}.

{\color{black}
High-resolution angle-resolved photoemission spectroscopy (ARPES) measurements were performed at the HR-branch of the I05-ARPES beamline \cite{hoesch2017facility} of Diamond Light Source. Single crystals of \UCG\ were cleaved \textit{in situ}, utilizing the top post method to expose pristine surfaces. Samples were cleaved at $T = 8$~K at a vacuum $<1 \times 10^{-10}$~mbar. All measurements were performed at $T = 8$~K with an energy resolution better than 12~meV and an angular resolution of 0.1$^{\circ}$, utilizing an MBS A1 hemispherical electron analyzer. The probing spot size was $<$50 microns.
Photon energies and light polarizations relevant to specific measurements are specified in the corresponding figure captions.}

\subsection{Theoretical}
Density functional theory (DFT) calculations were carried out using the pseudopotential projector-augmented wave method \cite{kresse1999ultrasoft} implemented in the Vienna Ab initio Simulation Package (VASP) \cite{kresse1993ab,kresse1996efficient}. An energy cutoff of 320~eV was used for the plane-wave basis set. Exchange-correlation effects were treated using the Perdew-Burke-Ernzerhof (PBE) generalized gradient approximation density functional \cite{perdew1996generalized}. An 11$\times$11$\times$7 $\Gamma$-centered k-point mesh was used to sample the Brillouin zone. Spin-orbit coupling effects were included self-consistently. The DFT+$U$ calculations utilized an effective Hubbard $U$ of 6~eV on the uranium 5$f$ or the lutetium 4$f$ states \cite{dudarev1998electron}. The \UCG\ unit cell was constructed using a simplified $P6/mmm$ cell with lattice parameters $a=a_\mathrm{m}$, $c=2c_\mathrm{m}$, where $a_\mathrm{m}$ and $c_\mathrm{m}$ are the lattice parameters of the average monoclinic cell discussed in $\S$\ref{sec:cryst}. A total energy tolerance of 10$^{-6}$~eV was used to determine the self-consistent charge density.

\section{Results and Discussion}
\subsection{Crystal structure and composition} \label{sec:cryst}
\UCG\ crystallizes in a modulated structure with the average unit cell having $C2/m$ space group symmetry. The lattice parameters of the average cell are $a=5.1680(3)$~\AA, $b=8.9509(5)$~\AA, $c=4.1452(2)$~\AA, and $\beta=90.023(2)^\circ$ (Fig.~\ref{fig:cell}), and the corresponding modulation vector is $\mathbf{q}=0.662(17)\mathbf{a^\ast}+0.501(8)\mathbf{c^\ast}$, within refinement error of a commensurate vector $\mathbf{q}=2/3\mathbf{a^\ast}+1/2\mathbf{c^\ast}$. Additional refinement details are provided in the Supplemental Material \cite{supplement}. 
{\color{black}The average {\color{black}monoclinic} cell is better described by the chemical formula U$_{0.5}$Cr$_3$Ge$_3$ (analogous to the Y$_{0.5}$Co$_{3}$Ge$_{3}$ structure \cite{buchholz1981intermetallische} {\color{black}with $Z=2$}), but for simplicity in comparisons to other 166 compounds, we will use \UCG\ as the formula unit (f.u.) for normalizations {\color{black}($Z=1$)}.

\begin{figure}
    \centering
    \includegraphics[width=0.9\columnwidth]{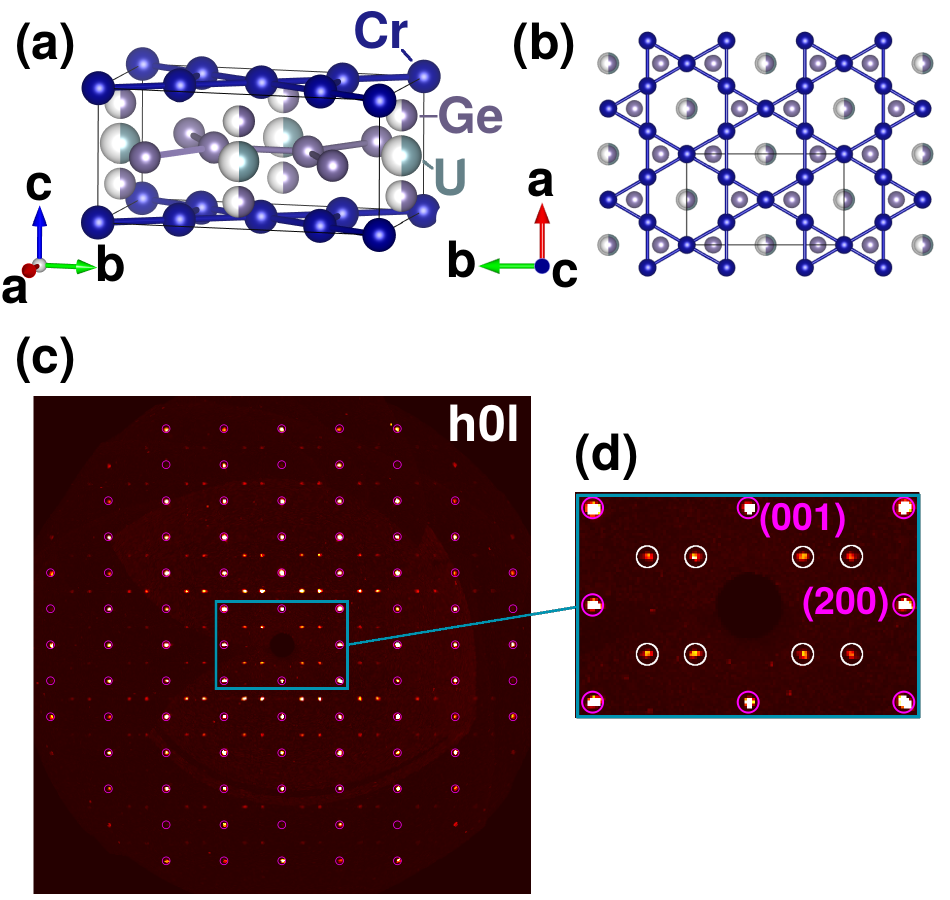}
    \caption{(a) Average unit cell of \UCG\ with half-occupancy uranium and germanium sites (b) View down the $c$ axis showing the kagome layer of chromium atoms (c) $h0l$ precession image for single crystal XRD on \UCG\ (d) Zoomed in view of the $h0l$ precession image showing the average cell reflections circled in pink and the modulation reflections at $l=\pm0.5$ circled in white}
    \label{fig:cell}
\end{figure}

For other stuffed CoSn materials, including UV$_6$Sn$_6$ \cite{patino2025incom,thomas2025uv6sn6}, modulated intergrowths of the ScFe$_6$Ga$_6$ ($Immm$) and ScFe$_6$Ge$_6$ ($P6/mmm$) structure types are common \cite{fredrickson2008origins}. However, the modulation behavior of \UCG\ is distinct from that of other 166s.}
{A 1/2$\mathbf{c^\ast}$ component for \UCG\ can be viewed as returning the {\color{black}average monoclinic} unit cell length along $c$ to that of the full 1-6-6 cell {\color{black}(ScFe$_6$Ge$_6$ type)}, rather than that of the half 0.5-3-3 cell {\color{black}(Y$_{0.5}$Co$_3$Ge$_3$ type)}, as is also the case for UV$_6$Sn$_6$ \cite{thomas2025uv6sn6}. A 2/3$\mathbf{a^\ast}$ modulation would triple the real-space $a$ axis.} 
The 2/3$\mathbf{a^\ast}$ (in-plane) component is distinct from that of other 166 intergrowths, for example  TbFeGe$_{3.5}$Ga$_{2.5}$ ($\mathbf{q}=2/3\mathbf{b^\ast}$) \cite{venturini2001crystallographic,venturini2006filling,fredrickson2008origins}, where the real-space propagation of the modulation in the kagome plane has a component perpendicular to that of \UCG, i.e. along $b$. 

\begin{figure}
    \centering
    \includegraphics[width=\columnwidth]{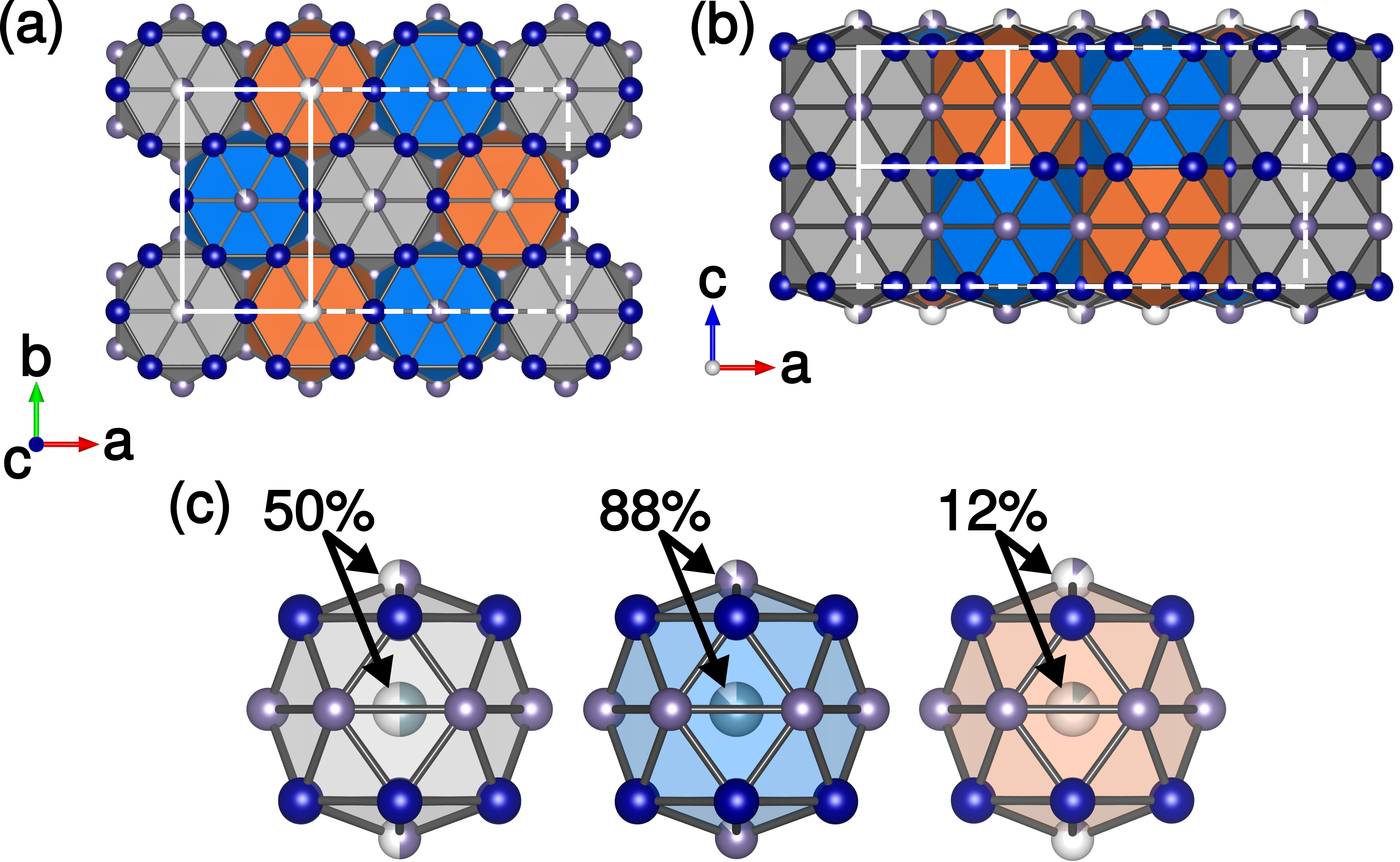}
    \caption{A 3$\times$1$\times$2 supercell that approximates the real-space modulation of \UCG\ is viewed down the average cell's (a) $c$ axis and (b) $b$ axis. Dashed and solid white lines indicate the bounds of the 3$\times$1$\times$2 supercell and the average cell, respectively. (c) The building blocks of the supercell are cages with half- (gray), \mbox{88\%-} (blue), and \mbox{12\%-} (orange) occupied U centers and Ge vertices. In the actual crystal, to maintain the correct stoichiometry and reasonable bond distances, the Ge vertices and the U center cannot be occupied in neighboring cages along $c$. Atom colors follow Fig.~\ref{fig:cell}.}
    \label{fig:supercell}
\end{figure}

To visualize the unit cell modulation, the modulation reflections [circled in white in Fig.~\ref{fig:cell}(c)] are refined with the average-cell reflections [circled in pink in Fig.~\ref{fig:cell}(c)]. The combined reflection set is reasonably modeled by a {\color{black}hexagonal super}cell with $P6_3/mmc$ space-group symmetry ($R_1=2.6\%$, $wR_2=8.7\%$). Since $P6_3/mmc$ is a supergroup of $C2/m$ \cite{ivantchev2000subgroupgraph}, the refined hexagonal {\color{black}super}cell can be transformed into a monoclinic {\color{black}super}cell with nominal $C2/m$ symmetry. The lattice vectors of the resulting monoclinic {\color{black}super}cell correspond to a $3\times1\times2$ supercell of the average monoclinic cell, and the periodicity matches a $\mathbf{q}=(2/3,0,1/2)$ modulation (Fig.~\ref{fig:supercell}), indicating it is an approximate representation of the modulation in real space. Additional details on the relationship between the $P6_3/mmc$ cell and the $C2/m$ cells are provided in the Supplemental Material \cite{supplement}. 

The {\color{black}monoclinic} supercell contains two unique, disordered U-Ge channels running along the $c$ axis. One contains half-occupied U and Ge sites (gray polyhedra), as in the Y$_{0.5}$Co$_{3}$Ge$_{3}$-type cell, while the other is split between primary [87.7(6)\%] and secondary [12.3(6)\%] U and Ge sites (blue and orange polyhedra), matching the SmMn$_6$Sn$_6$-type disorder \cite{malaman1997magnetic} observed in UNb$_6$Sn$_6$ \cite{riedel2025magnetic}. As established for the SmMn$_6$Sn$_6$-type structure, the primary and secondary U/Ge sites cannot be occupied in neighboring cages along $c$ in the actual crystal; otherwise, bond distances are unreasonable because of overlap between the positions of the top and bottom Ge vertices in neighboring cages. Further analysis may require 3+1 superspace group symmetry treatments, which are beyond the scope of this work.

\begin{figure}
    \centering
    \includegraphics[width=0.9\columnwidth]{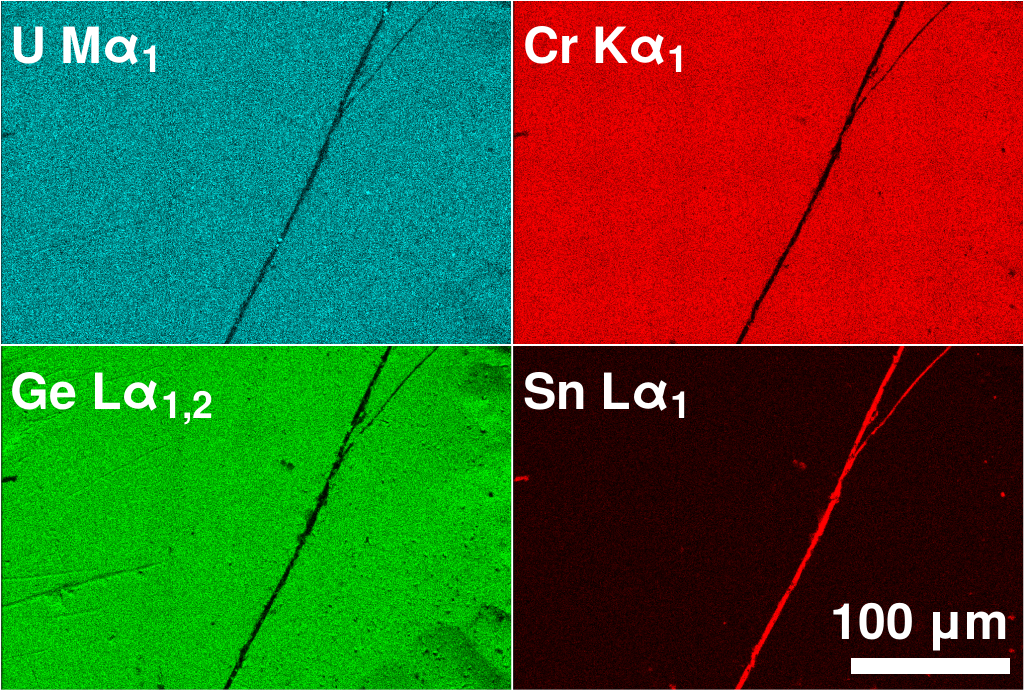}
    \caption{EDS maps of the surface of a polished \UCG\ crystal (black=no intensity). A Sn streak due to residual flux is apparent but localized. The bulk of the material is the expected 1:6:6 stoichiometry.}
    \label{fig:eds}
\end{figure}

To confirm the 1:6:6 stoichiometry is maintained, EDS data was collected on the polished surface of a crystal (Fig.~\ref{fig:eds}). The surface shows a uniform distribution of uranium, chromium, and germanium, giving an average U-Cr-Ge composition of 7.7(3)-46(2)-47(1)~at.\% over maps of four regions, which matches the expected 7.692-46.154-46.154~at.\% ratio. The pictured region also contains a streak of residual tin flux not fully removed by polishing. The isolation of tin to the streak demonstrates the lack of tin incorporation in the bulk.

Structural analysis was also performed for the reference compound \LCG. Single crystal diffraction did not show evidence of the site disorder reported for polycrystalline samples in Refs.~\cite{konyk2020electrical,romaka2024structure}. Instead, the data indicate a fully ordered $P6/mmm$ cell. The discrepancy may be due to the lower effective quenching temperature in our work (500$^\circ$C) than for the arc-melted samples in the previous works. Additional refinement details are in the Supplemental Material \cite{supplement}.

\subsection{Magnetic properties}
\begin{figure}
    \centering
    \subfloat{\includegraphics[width=0.5\columnwidth]{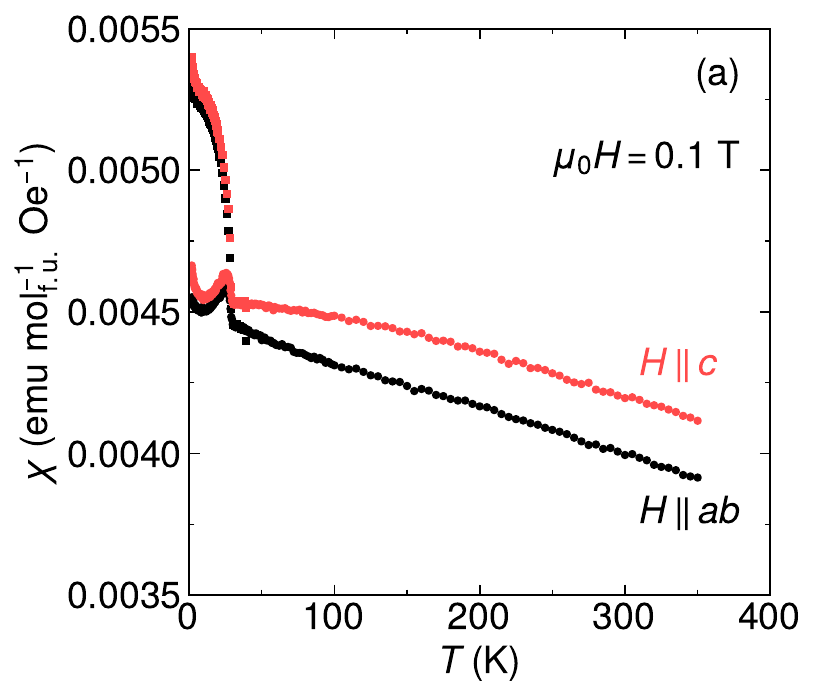}}    
    \subfloat{\includegraphics[width=0.46\columnwidth]{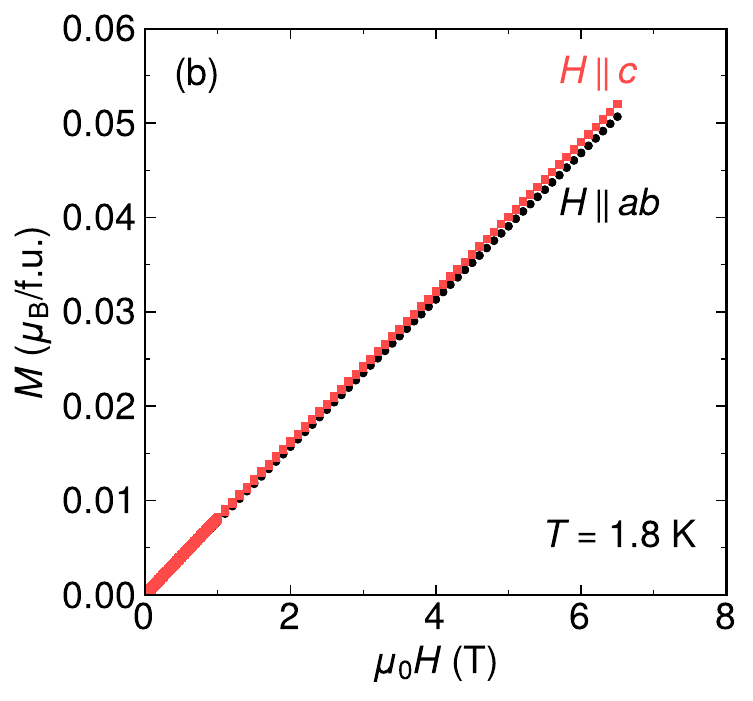}}
    \caption{(a) The zero-field-cooled and field-cooled magnetic susceptibilities of \UCG\ with field parallel to ($H{\parallel}ab$) and perpendicular to ($H{\parallel}c$) the kagome plane contain a ferro(ferri)magnetic transition at 30~K from an impurity. (b) The 1.8~K magnetization of \UCG\ is linear, featureless, and isotropic up to 6.5~T.}
    \label{fig:mag}
\end{figure}

The magnetic susceptibility, $\chi(T)$, of \UCG\  is small and {\color{black}nearly} linear from 350~K to 30~K, where a feature indicative of magnetic order appears [Fig.~\ref{fig:mag}(a)]. Magnetically ordered $R$Cr$_6$Ge$_6$ materials are ferrimagnets with $R$ and Cr magnetic sublattices, Cr moment magnitudes around 0.5~$\mu_\mathrm{B}$, and $R$ moments reduced from the free-ion values \cite{schobinger1997atomic,schobinger1997ferrimagnetism}. For the magnetically ordered lanthanide systems, a more even split in lanthanide occupation of the $z=0$ and $z=c/2$ lattice sites correlates with a higher magnetic ordering temperature \cite{schobinger1997ferrimagnetism}. \UCG\ represents the maximum limit of site disorder in that framework, where the $z=0$ and $z=c/2$ lattice sites are equally occupied, leading to a halving of the unit cell (ignoring the monoclinic modulation). A transition temperature higher than that of TbCr$_6$Ge$_6$ ($T_\mathrm{C}=10.3$~K \cite{schobinger1997atomic}) is, therefore, reasonable. However, the magnetic susceptibility of \UCG\ does not show Curie-Weiss behavior in the paramagnetic region. Though fits to the relationship $\chi^{-1}=(T-\theta_\mathrm{CW})/C$ give an effective moment of 8.84~$\mu_\mathrm{B}$/f.u. ($H{\parallel}ab$) or 9.49~$\mu_\mathrm{B}$/f.u. ($H{\parallel}c$), and Curie-Weiss temperatures ($\theta_\mathrm{CW}$) near \mbox{-2000}~K, {\color{black}these values are likely unphysical}. 
The shape and magnitude of $\chi$, therefore, suggest that the 30~K transition is due to an impurity, and \UCG\ is a Pauli paramagnet. 
In contrast, UV$_6$Sn$_6$ and UNb$_6$Sn$_6$, where the transition metals do not carry a magnetic moment, exhibit localized 5$f$ electron magnetism \cite{patino2025incom,thomas2025uv6sn6,riedel2025magnetic}.

A review of possible Cr--Ge and U--Ge binary impurity phases finds no compounds with a transition near 30~K \cite{zagryazhskii1968magnetic,kolenda1980esca,sato1983magnetic,sato1988electrical,onuki1992magnetic,troc2002magnetotransport,ghimire2012complex,ji2025small}, and there are no reported U--Cr binary compounds \cite{venkatraman1985CrU}.
A possible ternary impurity phase is U$_3$CrGe$_5$, which has a reported ferromagnetic transition near 25-30~K, though that transition may also be extrinsic \cite{moussa2019overview}. {\color{black}Notably, the magnitude of the susceptibility jump varied significantly between \UCG\ batches (up to 4$\times$ larger than in Fig.~\ref{fig:mag}), as did the anisotropic $\chi$($T$) high-temperature data and the $M$($H$) low-field data (see the Supplemental Material \cite{supplement}). Therefore, we assume that the magnetic susceptibility jump in Ref.~\cite{moussa2019overview} and the feature in Fig.~\ref{fig:mag}(a) are from the same magnetic impurity.}

If one assumes that the U $5f$ electrons in \UCG\ are itinerant and their Pauli paramagnetism contributes to the magnetic susceptibility, the estimated Pauli susceptibility is $\chi_0 \sim1.2\times10^{-3}$ emu/mol, given a Wilson ratio $R_W$ [\,=($\pi^2 k_B^2/\mu_{eff})(\chi_0/\gamma$)\,] of 1 {\color{black}for a noninteracting electron gas} and the measured Sommerfeld coefficient $\gamma=86.5$ mJ/mol-K$^2$ discussed below.  This $\chi_0$ is $\sim$3.5$\times$ smaller than the measured value. If this interpretation is correct, then either there may be U $5f$ ferromagnetic correlations, or a paramagnetic contribution from the Cr $d$-electrons may dominate the magnetic susceptibility. 

An alternative explanation for the small magnetic susceptibility of  \UCG\ is that the crystalline electric field (CEF) splits a $J=4$ ($5f^2$ valence) multiplet into 9 singlets, with a small van Vleck contribution to $\chi(T)$ due to a large splitting between the nonmagnetic singlet ground state to another singlet first-excited state (perhaps with additional contributions to $\chi(T)$ from the Cr $d$-electrons). The Cr--Ge cages of \UCG\ are smaller than the V--Sn and Nb--Sn cages in UV$_6$Sn$_6$ and UNb$_6$Sn$_6$. The uranium atoms that fill the Cr--Ge cages may, therefore, be more stable in a 5$f^2$ valence configuration. The large number of symmetry-allowed crystal field parameters for the monoclinic structure means that crystal field fits to $\chi(T)$ and $M(H)$ are underconstrained. Therefore, further measurements, e.g., resonant inelastic x-ray scattering, are necessary to understand the valence state {\color{black}and crystal field levels} of the U $5f$ electrons in  \UCG. 

The magnetization, $M(H)$, of \UCG\ was measured at 1.8~K with the applied magnetic field parallel or perpendicular to the kagome lattice plane [Fig.~\ref{fig:mag}(b)]. 
There is no appreciable anisotropy, and the moment size is only 0.052~$\mu_\mathrm{B}$/f.u. at 6.5~T. 
Other $R$Cr$_6$Ge$_6$ compounds approach saturation near 7~T with magnetization around 1--10~$\mu_\mathrm{B}$/f.u. \cite{brabers1994magnetic,yang2024crystal}. A small curvature below 0.5~T and a remnant magnetization of order 10$^{-3}$~$\mu_\mathrm{B}$/f.u. were observed in some crystals of \UCG\ where the impurity transition magnitude in $\chi$ was also larger {\color{black}(Fig.~S2 \cite{supplement})}, providing further evidence of a small ferromagnetic impurity leading to the 30~K feature. As with $\chi$, the small and isotropic magnetization suggests itinerant uranium 5$f$ electrons and paramagnetism. Similarly small values of the magnetization are observed for several $R$Fe$_6$Ge$_6$ compounds with nonmagnetic $R$ elements, but in each, Fe orders well above room temperature \cite{mazet2001macroscopic}. No decrease in $\chi$ with decreasing temperature is observed for \UCG, suggesting there is not an analogous high-temperature ordering transition of the chromium sublattice.

\subsection{Heat capacity and band structure}
\begin{figure}
    \centering
    \includegraphics[width=0.9\columnwidth]{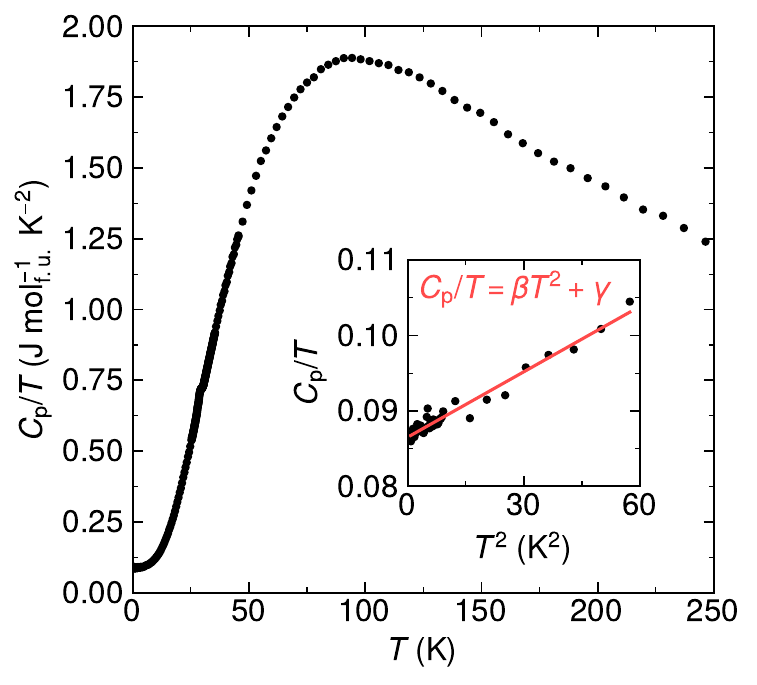}
    \caption{The heat capacity ($C_\mathrm{p}/T$) of \UCG\ is shown with (inset) a Sommerfeld coefficient ($\gamma$) fit to the low-temperature heat capacity, giving 86.5~mJ~mol$^{-1}$~K$^{-2}$.}
    \label{fig:hc}
\end{figure}

The heat capacity $C_\mathrm{p}(T)$ of \UCG\, plotted as $C/T$ (Fig.~\ref{fig:hc}), is mostly featureless but exhibits two bumps near 30~K attributed to an impurity phase (Fig.~S3 \cite{supplement}). A fit of $C/T={\beta}T^2+\gamma$ between $0.7{\leq}T^2{\leq}60$~K yields $\gamma = 86.5(1)$~mJ~mol$^{-1}$~K$^{-2}$ and $\beta = 0.289(8)$~mJ~mol$^{-1}$~K$^{-4}$. The large Sommerfeld coefficient of \UCG\ is comparable to that reported for LuFe$_6$Ge$_6$, YbFe$_6$Ge$_6$, and Lu$_{1-x}$Fe$_6$Sn$_6$ ($\gamma=87-90$~mJ~mol$^{-1}$~K$^{-2}$ \cite{avila2005direct,lyu2024anomalous,shi2025weak}), though smaller than that of YbV$_6$Sn$_6$ ($\gamma\sim400$~mJ~mol$^{-1}$~K$^{-2}$ \cite{guo2023triangular}) and YbMn$_6$Sn$_6$ ($\gamma\sim144$~mJ~mol$^{-1}$~K$^{-2}$ \cite{lou2025strongly}).
Additionally, the Debye temperature of \UCG\ is determined to be ${\theta}_{\mathrm{D}}= [(12{\pi}^4 n R)/({5\beta})]^{1/3}$=444(4)~K, where $n$ is the number of atoms per formula unit (13) and $R$ is the gas constant. 
${\theta}_{\mathrm{D}}$ is consistent with the larger values (calculated with DFT) for $R$Cr$_6$Ge$_6$ compounds with lighter $R$ elements \cite{romaka2024structure}. 
As a comparison, the heat capacity of \LCG\ was also collected (Fig.~S4(a) \cite{supplement}). The fit Sommerfeld coefficient and Debye temperature were $\gamma=57.3(3)$~mJ~mol$^{-1}$~K$^{-2}$ and ${\theta}_{\mathrm{D}}=458(9)$~K for \LCG. The Sommerfeld coefficient of \LCG\ is similar to that of YCr$_6$Ge$_6$ (66~mJ~mol$^{-1}$~K$^{-2}$), which has electronic bands near $E_\mathrm{F}$ associated with the chromium kagome lattice \cite{ishii2013ycr6ge6}. The similarity suggests that the larger electronic heat capacity of \UCG\ may be driven by band structure features associated with not only chromium but also additional uranium 5$f$ states that increase the DOS near $E_\mathrm{F}$. 

\begin{figure*}
    \centering
    \subfloat{\includegraphics[width=\columnwidth]{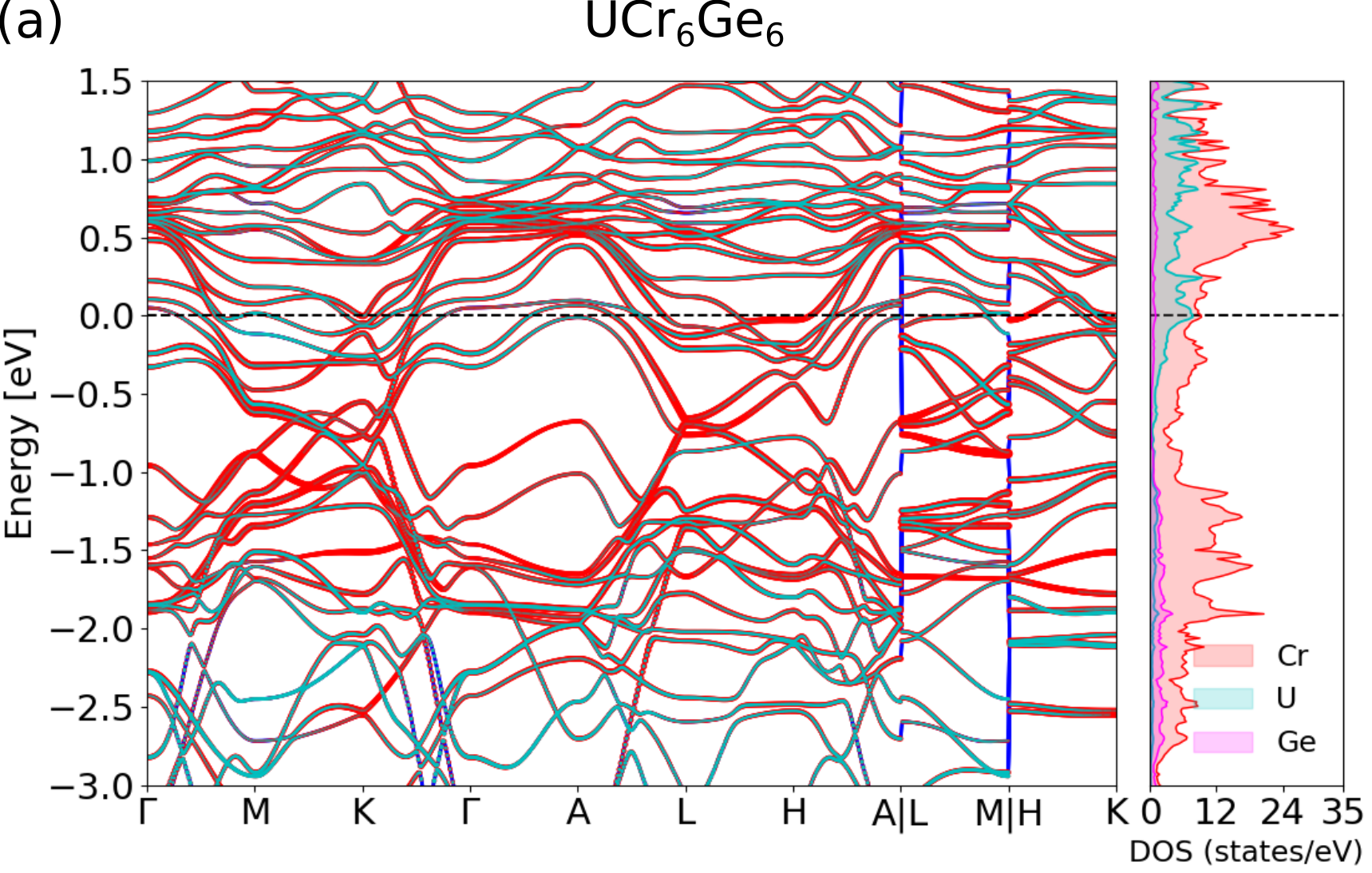}}
    \hspace{1em}
    \subfloat{\includegraphics[width=\columnwidth]{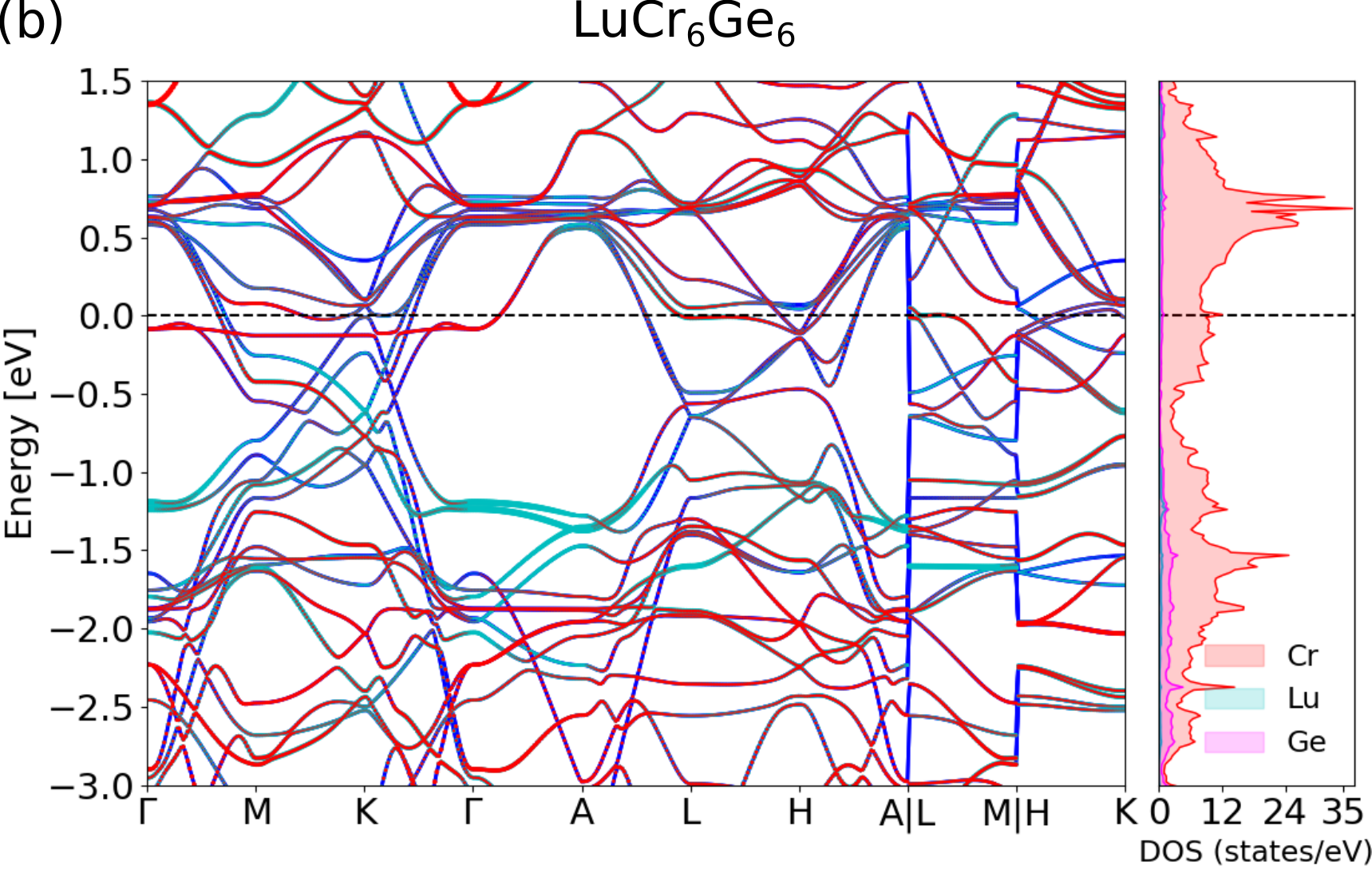}}
    \caption{The electronic band structure of (a) \UCG\ is calculated with uranium 5$f$ electrons in the valence band, and the band structure of (b) \LCG\ is calculated for comparison. The band color and thickness indicate U/Lu (teal), Cr (red), and Ge (pink) weight.}
    \label{fig:bands}
\end{figure*}

To assess whether the electronic heat capacity is influenced by chromium kagome flatbands and uranium 5$f$ states, the electronic band structure was calculated for an ordered \UCG\ cell with uranium 5$f$ electrons in the valence [Fig.~\ref{fig:bands}(a)] and for \LCG\ [Fig.~\ref{fig:bands}(b)]. Additional band structures for each material with an effective Hubbard $U$ are provided in the Supplemental Material \cite{supplement}. The calculated bands are shown with colors indicating weight from uranium/lutetium (teal), chromium (red), or germanium (pink) states. The general features of the \UCG\ band structure are, unsurprisingly, similar to those reported for other 166s \cite{thomas2025uv6sn6,riedel2025magnetic,hu2022tunable,ortiz2024stability,kim2023infrared}, but for \UCG, flatbands typically associated with the kagome lattice along the $M$--$K$ path sit below $E_\mathrm{F}$. The relative $E_\mathrm{F}$ shift also places a flat valence band with hybrid uranium and chromium character along the $\Gamma$--$A$ Brillouin zone path close to $E_\mathrm{F}$, further demonstrating the band tuning possibilities provided by expanding the 166 family to 5$f$ systems. 

Comparing the \UCG, \LCG, and previously published YCr$_6$Ge$_6$ \cite{ishii2013ycr6ge6} band structures reveals that the $R$Cr$_6$Ge$_6$ structures are conducive to a significant DOS at $E_\mathrm{F}$. The total DOS at $E_\mathrm{F}$ for a \UCG\ formula unit is 21.64~states/eV (13.79~states/eV for \LCG). 
The equation \[\frac{C_\mathrm{el}}{T}=\gamma=\frac{\pi^2}{3}~k_\mathrm{B}^2~D(E_\mathrm{F})\] relates the Sommerfeld coefficient to the DOS at $E_\mathrm{F}$ [$D(E_\mathrm{F})$] and the Boltzmann constant ($k_\mathrm{B}$). 
The predicted Sommerfeld coefficient is 51.0~mJ~mol$^{-1}$~K$^{-2}$ for \UCG\ and 32.5~mJ~mol$^{-1}$~K$^{-2}$ for \LCG. The ratio of the experimental to predicted value is, therefore, 1.7 for \UCG\ and 1.8 for \LCG, indicating moderate enhancement. 
When an effective Hubbard $U$ is included in the band structure calculation to localize the uranium 5$f$ electrons, the chromium DOS near $E_\mathrm{F}$ is mostly unchanged, and the uranium weight at $E_\mathrm{F}$ decreases to nearly zero (Fig.~S5(a) \cite{supplement}), resulting in a predicted Sommerfeld coefficient of 26.4~mJ~mol$^{-1}$~K$^{-2}$.

{\color{black}
\subsection{ARPES}
\begin{figure*}[]
    \centering
    \subfloat{\includegraphics[width=1.5\columnwidth]{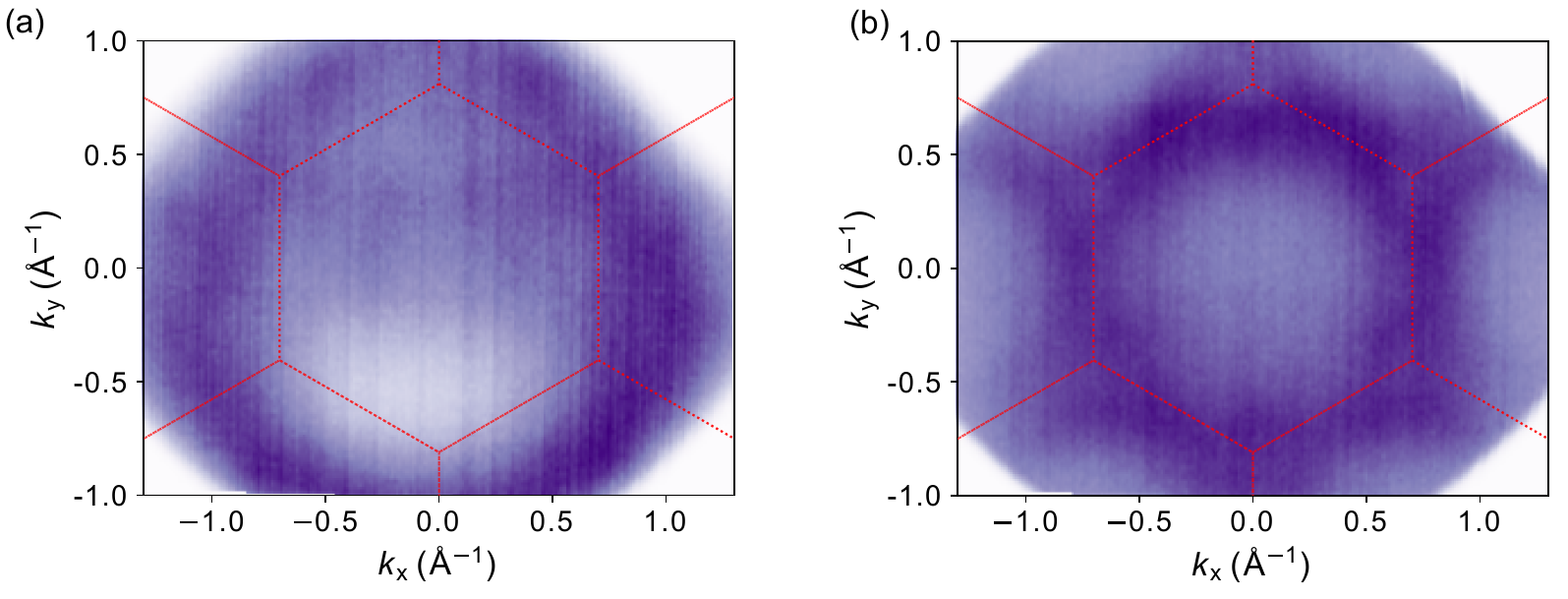}}   
    \caption{$k_\mathrm{x}–k_\mathrm{y}$ constant-energy surfaces at energies of (a) $E-E_\mathrm{F} =0 \pm 0.02$~eV and (b) $E-E_\mathrm{F}=-0.3 \pm 0.02$~eV. At the Fermi level (a) there is little structure to the constant energy surface while at (b) $E-E_\mathrm{F}=-0.3$~eV, the more typical kagome Fermi surface is recovered. The Brillouin zones are delineated by the dashed red lines. ($h\nu=98$~eV, LH polarization, $T = 8$~K)}
    \label{fig:kxkyconstantE}
\end{figure*}
High-resolution ARPES was used to probe the electronic structure of \UCG\ directly to determine if the enhanced Sommerfeld coefficient observed is related to the presence of flat bands in the vicinity of the Fermi level. 
Constant energy contours in the $k_\mathrm{x}$-$k_\mathrm{y}$ plane were measured and are presented in Fig.~\ref{fig:kxkyconstantE}. As shown in Fig.~\ref{fig:kxkyconstantE}(a), there is little structure at the Fermi surface within the first Brillouin zone, while spectral weight with little momentum variation is observed in neighboring zones. This is in sharp contrast to the Fermi surface observed in UV$_6$Sn$_6$, in which the prototypical six-fold symmetric kagome structure is clearly resolved \cite{thomas2025uv6sn6}. However, a six-fold symmetric structure as well as triangular pockets about the $K$ point are observed in a constant energy contour recorded at $E-E_\mathrm{F} =-0.3 \pm 0.02$~eV, as displayed in Fig.~\ref{fig:kxkyconstantE}(b). This suggests an electron band filling which shifts the Fermi level of order 300~meV in \UCG\ relative to UV$_6$Sn$_6$. 

To search for evidence of 3$d$ Cr kagome flatbands near the Fermi level, valence band spectra were collected with photon energies of 92~eV (uranium Cooper minimum) and 108~eV (uranium 5$d$-5$f$, or O-edge, resonance condition), utilizing linear horizontal (LH) and linear vertical (LV) polarized light to account for matrix element effects. These energies correspond to $k_\mathrm{z}$ values that lie close to the $L-A-L$ and the $M-\Gamma-M$ planes, respectively (Fig.~S6 \cite{supplement}). 
When probed with LH polarization, a clear flatband is observed near the Fermi level [Fig.~\ref{fig:dispersions}(a)], the spectral weight of which is suppressed in the first Brillouin zone due to matrix element effects. With LV-polarized light, the band is also weakly observed in the first Brillouin zone [Fig.~\ref{fig:dispersions}(c)].
As this measurement is performed at the Cooper minimum of uranium, we ascribe Cr character, likely from even-parity Cr $3d_{z^2}$ orbitals, to these flat bands. The presence of a kagome flat band in the vicinity of the Fermi level is a strong candidate for the origin of the enhanced Sommerfeld coefficient observed in \UCG, contrasting with UV$_6$Sn$_6$, where a flat band near the Fermi level is absent. 

Furthermore, in Fig.~\ref{fig:dispersions}(b,d), we show valence band dispersions that enhance uranium character due to the uranium $5d$-$5f$ resonance condition. In Fig.~\ref{fig:dispersions}(b,d), there is enhanced momentum-independent spectral weight at the Fermi level characteristic of conduction--$f$-electron ($c-f$) hybridization in heavy fermion systems. The larger spectral weight found in \UCG\ compared to that of UV$_6$Sn$_6$ indicates that the $5f$ electrons in \UCG\ are more itinerant than in UV$_6$Sn$_6$. These results are consistent with the scenario of Pauli paramagnetism in \UCG\ arising from itinerant $5f$ electrons, while the more localized $5f$ electrons in UV$_6$Sn$_6$ lead to magnetic order \cite{thomas2025uv6sn6,patino2025incom}.

\begin{figure*}[]
    \centering
    \subfloat{\includegraphics[width=1.5\columnwidth]{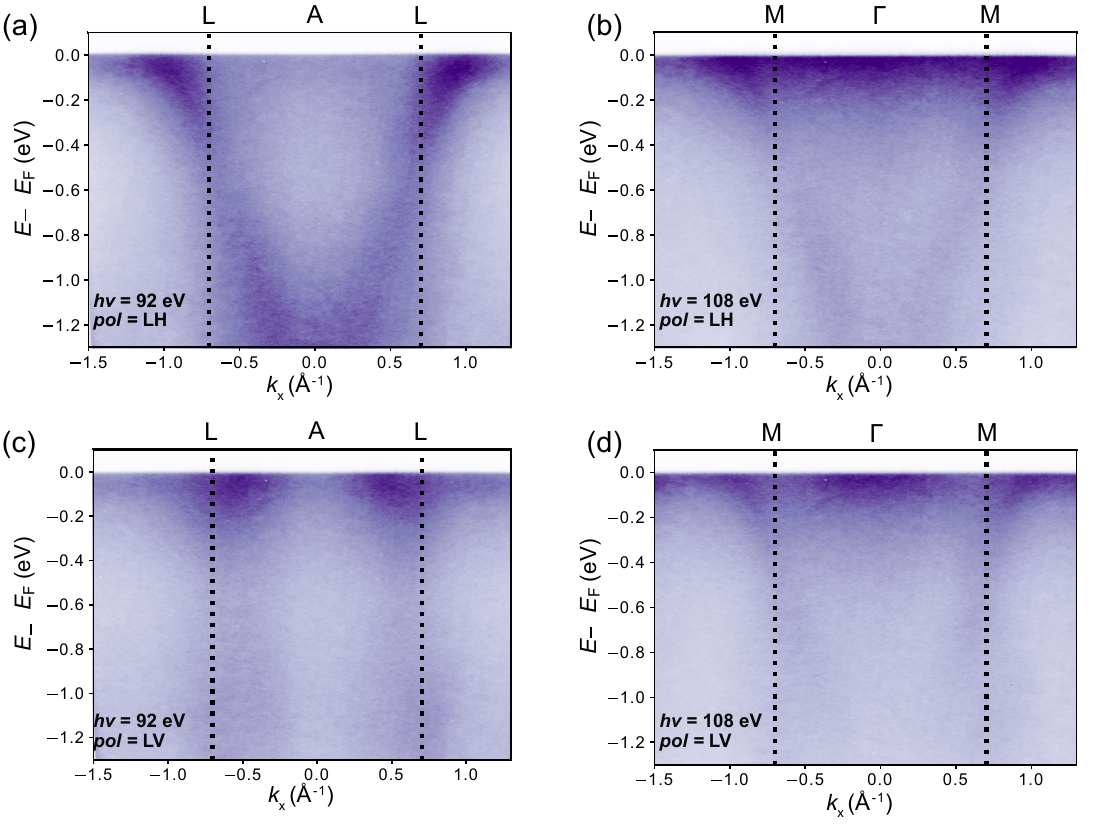}}   
    \caption{Valence band dispersions measured close to the (a,c) $L-A-L$ plane and (b,d) $M-\Gamma-M$ plane. Dashed black lines define the zone boundaries. (a,c) utilize the U Cooper minimum to strongly suppress the U photoemission cross section, while (b,d) are on-resonance (U $5d$-$5f$), enhancing U weight near the Fermi level.}
    \label{fig:dispersions}
\end{figure*}
}

\subsection{Resistivity}
\begin{figure}[]
    \centering
    \subfloat{\includegraphics[width=0.49\columnwidth]{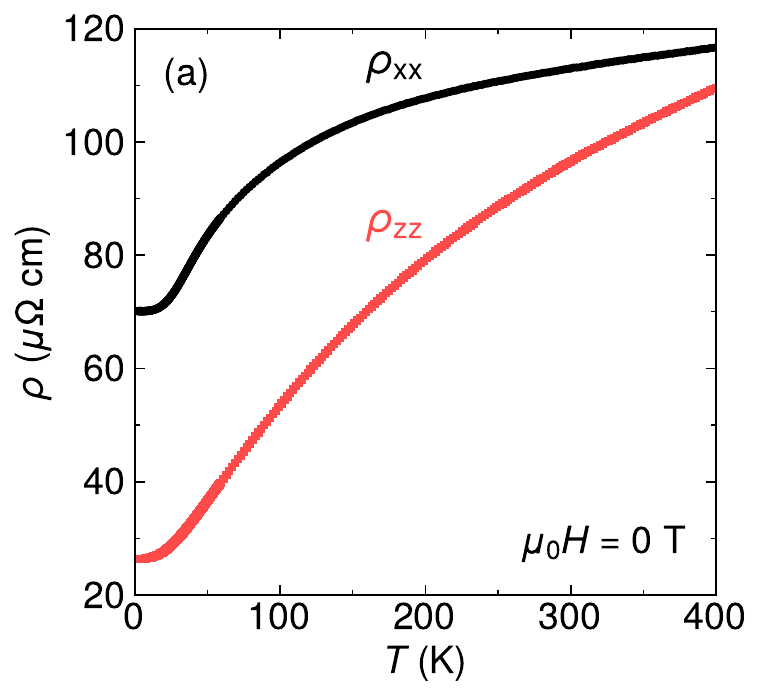}}    \subfloat{\includegraphics[width=0.465\columnwidth]{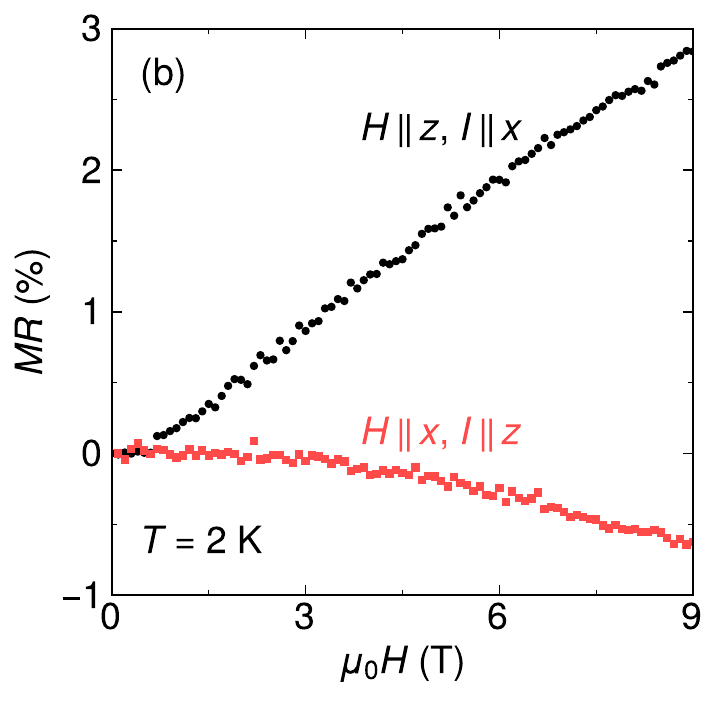}}
    \caption{(a) The zero-field longitudinal resistivity of \UCG\ collected with current applied in the kagome plane ($\rho_\mathrm{xx}$) and perpendicular to it ($\rho_\mathrm{zz}$) is shown along with (b) the 2~K magnetoresistance for each orientation with a magnetic field applied perpendicular to the current direction.}
    \label{fig:rho_MR}
\end{figure}
Figure~\ref{fig:rho_MR}(a) shows the zero-field longitudinal resistivity of \UCG\ with current applied in the kagome plane ($\rho_\mathrm{xx}$) and perpendicular to it ($\rho_\mathrm{zz}$). The residual resistivity ratio [$RRR=\rho$(300~K)/$\rho$(1.8~K)] is 1.6 for $\rho_\mathrm{xx}$ and 3.7 for $\rho_\mathrm{zz}$. Both are lower than the reported ratios for other $R$Cr$_6$Ge$_6$ crystals, the structures of which are less disordered \cite{konyk2020electrical,romaka2024structure}, but the values are similar to those of UNb$_6$Sn$_6$ (3.3 \cite{riedel2025magnetic}) and UV$_6$Sn$_6$ (2-3 \cite{patino2025incom,thomas2025uv6sn6}). Neither orientation's resistivity contains a feature near 30~K, and the first derivative of both orientations only shows a broad hump at the inflection point of the $\rho(T)$ curve ({\color{black}Fig.~S7} \cite{supplement}). The resistivity with current along the $c$ axis ($\rho_\mathrm{zz}$) is lower than the resistivity with current within the $ab$ plane ($\rho_\mathrm{xx}$), as is also observed for YCr$_6$Ge$_6$ \cite{ishii2013ycr6ge6} and for several CoSn-structure compounds \cite{meier2020flat}. Figure~\ref{fig:rho_MR}(b) shows the magnetoresistance (MR) up to 9~T for both current directions with a perpendicular applied magnetic field. 
The magnetoresistance was calculated using
\[MR = \frac{\rho(H)-\rho(H=0)}{\rho(H=0)}\times100\%.\]
As with the zero-field, temperature-dependent resistivity, no transitions appear. The changes are small and monotonic for both orientations and are consistent with preferred conduction perpendicular to the kagome lattice. {\color{black}Additional magnetoresistance ($H{\parallel}x$, $I{\parallel}z$) and Hall resistivity ($\rho_\mathrm{yx}$) data at 2, 10, 20, 30, and 50~K are presented in the Supplemental Material \cite{supplement}. At higher temperatures, the magnetoresistance is approximately flat and near 0\%, and the Hall resistivity indicates predominant electron carriers in a one-band approximation. A linear fit of $\rho_\mathrm{yx}(H)$ to $\rho_\mathrm{yx}=R_H\mu_\mathrm{0}H$ at 2~K results in a carrier concentration $n_e = 1/(R_Hq) = 4.27 \times 10^{23}$ cm$^{-3}$ (Fig.~S9), where $R_H$ is the Hall coefficient and $q$ is the electron charge.}

\section{Conclusions}
\UCG\ crystallizes in a structure with an average Y$_{0.5}$Co$_3$Ge$_3$-type unit cell and a modulation that lowers the space-group symmetry to monoclinic. The modulation [$\mathbf{q}=0.662(17)\mathbf{a^\ast}+0.501(8)\mathbf{c^\ast}$] of \UCG\  is distinct from other 166 materials in that it contains an $\mathbf{a^\ast}$ component and no $\mathbf{b^\ast}$ component. The isotropic temperature and magnetic field dependence of the magnetization contrast with other $R$Cr$_6$Ge$_6$ and uranium-containing 166 materials with localized magnetic moments. Instead, the shape of the high-temperature susceptibility and the small magnetization strongly suggest that \UCG\ is Pauli paramagnetic, though a ground state singlet due to a CEF splitting of a U 5$f^2$ multiplet cannot be ruled out. A transition at 30~K in the magnetic susceptibility and heat capacity are attributed to an impurity phase. 

The electronic heat capacity, quantified by a Sommerfeld coefficient of 86.5~mJ~mol$^{-1}$~K$^{-2}$, is moderately enhanced in \UCG, possibly due to flatbands near the Fermi level. Electronic band structure calculations suggest that the kagome flatbands and uranium states may increase the DOS near $E_\mathrm{F}$. 
{\color{black}The ARPES study reveals that the Fermi level is shifted $\sim$300~meV in \UCG\ relative to UV$_6$Sn$_6$. This shift results in Cr 3$d$ flatbands now residing near the Fermi level, giving a large density of states from which the enhanced Sommerfeld coefficient can be explained. Furthermore, the U $5f$ states exhibit $c-f$ hybridization, indicating more itinerant $5f$ character in \UCG, consistent with a Pauli paramagnetic state.}
The anisotropic resistivity and magnetoresistance reveal a preference for conduction perpendicular to the kagome lattice, consistent with other CoSn-type and stuffed-CoSn-type materials. The moderately enhanced Sommerfeld coefficient, isotropic magnetic properties, and unique structural modulation distinguish \UCG\ from other uranium-containing 166 compounds and demonstrate the promise of tuning electronic band structure features and magnetic ground states in 5$f$ 166 materials.
\\
\begin{acknowledgments}
Crystal growth, physical property measurements, and magnetic property measurements were supported by the U.S. Department of Energy, Office of Basic Energy Sciences, Division of Materials Science and Engineering project ``Quantum fluctuations in narrow band systems." Structural characterization was also supported by the Laboratory Directed Research and Development program. C.S.K. {\color{black}and P.A.E.M.} gratefully acknowledges the support through the LANL/LDRD Program and the G. T. Seaborg Institute. 
{\color{black}We gratefully acknowledge Diamond Light Source for access to Beamline I05 (Proposal SI41855-1), which contributed to the results presented here.}
Scanning electron microscope and energy dispersive X-ray measurements were performed at the Electron Microscopy Lab and supported by the Center for Integrated Nanotechnologies, an Office of Science User Facility operated for the U.S. Department of Energy Office of Science. The electronic structure calculations were supported in part by the Center for Integrated Nanotechnologies, a DOE BES user facility, in partnership with the LANL Institutional Computing Program for computational resources. Additional computations were performed at the National Energy Research Scientific Computing Center (NERSC), a U.S. Department of Energy Office of Science User Facility located at Lawrence Berkeley National Laboratory, operated under Contract No. DE-AC02-05CH11231 using NERSC award ERCAP0028014.

\end{acknowledgments}

\bibliography{UCr6Ge6.bib}

@misc{supplement,
    key = "",
    note = "See Supplemental Material at [URL will be inserted by publisher] for additional information, including Refs.~\cite{malaman1997magnetic,ivantchev2000subgroupgraph}."
}

@article{fredrickson2008origins,
  title={Origins of superstructure ordering and incommensurability in stuffed {CoSn}-type phases},
  author={Fredrickson, Daniel C and Lidin, Sven and Venturini, Gerard and Malaman, Bernard and Christensen, Jeppe},
  journal={Journal of the American Chemical Society},
  volume={130},
  number={26},
  pages={8195--8214},
  year={2008},
  publisher={ACS Publications}
}

@article{thomas2025uv6sn6,
    author={Thomas, S M and Kengle, C S and Simeth, W and Lim, Chan-young and Riedel, Z W and Allen, K and Schmidt, A and Ruf, M and Gim, Seonggeon and Thompson, J D and Ronning, F and Scheie, A O and Lane, C and Denlinger, J D and Blanco-Canosa, S and Zhu, Jian-Xin and Bauer, E D and Rosa, P F S},
    title={Unusual $5f$ magnetism in new kagome material {UV$_6$Sn$_6$}},
    journal={npj quantum materials},
    year={2025},
    volume={10},
    pages={66}
}

@article{patino2025incom,
  title={Incommensurate and commensurate antiferromagnetic orders in the kagome compound {UV$_6$Sn$_6$}},
  author={Amano Patino, Midori and Raymond, Stephane and Knebel, Georg and Le Berre, Pierre and Savvin, Stanislav and Pachoud, Elise and Ressouche, Eric and Fettinger, James C and Leynaud, Olivier and Pecaut, Jacques and Klavins, Peter and Hasselbach, Klaus and Brison, Jean-Pascal and Lapertot, Gerard and Taufour, Valentin},
  journal={Physical Review B},
  volume={111},
  number={17},
  pages={174432},
  year={2025},
  publisher={APS}
}

@article{avila2005direct,
  title={Direct observation of {Fe} spin reorientation in single-crystalline {YbFe$_6$Ge$_6$}},
  author={Avila, M A and Takabatake, T and Takahashi, Y and Bud’ko, S L and Canfield, P C},
  journal={Journal of Physics: Condensed Matter},
  volume={17},
  number={43},
  pages={6969},
  year={2005},
  publisher={IOP Publishing}
}

@article{lyu2024anomalous,
  title={Anomalous {Hall} effect and electronic correlation in a spin-reoriented kagome antiferromagnet {LuFe$_6$Sn$_6$}},
  author={Lyu, Meng and Liu, Yang and Zhang, Shen and Liu, Junyan and Yang, Jinying and Wang, Yibo and Feng, Yiting and Dong, Xuebin and Wang, Binbin and Wei, Hongxiang and Liu, Enke},
  journal={Chinese Physics B},
  volume={33},
  number={10},
  pages={107507},
  year={2024},
  publisher={IOP Publishing}
}

@article{yang2024crystal,
  title={Crystal growth, magnetic and electrical transport properties of the kagome magnet {$R$Cr$_6$Ge$_6$} ({$R$= Gd--Tm)}},
  author={Yang, Xingyu and Zeng, Qingqi and He, Miao and Xu, Xitong and Du, Haifeng and Qu, Zhe},
  journal={Chinese Physics B},
  volume={33},
  number={7},
  pages={077501},
  year={2024},
  publisher={IOP Publishing}
}

@article{romaka2024structure,
  title={Structure, bonding, and properties of {$R$Cr$_6$Ge$_6$} intermetallics ({$R$=Gd--Lu})},
  author={Romaka, V V and Romaka, L and Konyk, M and Corredor, L T and Srowik, K and Kuzhel, B and Stadnyk, Yu and Yatskiv, Yu},
  journal={Journal of Solid State Chemistry},
  volume={338},
  pages={124874},
  year={2024},
  publisher={Elsevier}
}

@article{riedel2025magnetic,
  title={Magnetic order and physical properties of the kagome metal {UNb$_6$Sn$_6$}},
  author={Riedel, Z W and Simeth, W and Kengle, C S and Thomas, S M and Thompson, J D and Scheie, A O and Ronning, F and Lane, C and Zhu, Jian-Xin and Rosa, P F S and Bauer, E D},
  journal={Physical Review Materials},
  volume={9},
  pages={084401},
  year={2025},
  publisher={APS}
}

@article{mazet2001macroscopic,
  title={Macroscopic magnetic properties of the {HfFe$_6$Ge$_6$}-type {RFe$_6$X$_6$} ({X= Ge or Sn}) compounds involving a non-magnetic {R} metal},
  author={Mazet, T and Malaman, B},
  journal={Journal of Alloys and Compounds},
  volume={325},
  number={1-2},
  pages={67--72},
  year={2001},
  publisher={Elsevier}
}

@article{hu2022tunable,
  title={Tunable topological {Dirac} surface states and van {Hove} singularities in kagome metal {GdV$_6$Sn$_6$}},
  author={Hu, Yong and Wu, Xianxin and Yang, Yongqi and Gao, Shunye and Plumb, Nicholas C and Schnyder, Andreas P and Xie, Weiwei and Ma, Junzhang and Shi, Ming},
  journal={Science Advances},
  volume={8},
  number={38},
  pages={eadd2024},
  year={2022},
  publisher={American Association for the Advancement of Science}
}

@article{ortiz2024stability,
  title={Stability Frontiers in the {AM$_6$X$_6$} Kagome Metals: The {LnNb$_6$Sn$_6$ (Ln:Ce-Lu,Y)} Family and Density-Wave Transition in {LuNb$_6$Sn$_6$}},
  author={Ortiz, Brenden R and Meier, William R and Pokharel, Ganesh and Chamorro, Juan and Yang, Fazhi and Mozaffari, Shirin and Thaler, Alex and Gomez Alvarado, Steven J and Zhang, Heda and Parker, David S and Samolyuk, German D and Paddison, Joseph A M and Yan, Jiaqiang and Ye, Feng and Sarker, Suchismita and Wilson, Stephen D and Miao, Hu and Mandrus, David and McGuire, Michael A},
  journal={Journal of the American Chemical Society},
  year={2024},
  publisher={ACS Publications}
}

@article{ishii2013ycr6ge6,
  title={{YCr$_6$Ge$_6$} as a candidate compound for a kagome metal},
  author={Ishii, Yui and Harima, Hisatomo and Okamoto, Yoshihiko and Yamaura, Jun-ichi and Hiroi, Zenji},
  journal={Journal of the Physical Society of Japan},
  volume={82},
  number={2},
  pages={023705},
  year={2013},
  publisher={The Physical Society of Japan}
}

@article{kim2023infrared,
  title={Infrared probe of the charge density wave gap in {ScV$_6$Sn$_6$}},
  author={Kim, Dong Wook and Liu, Shuyuan and Wang, Chongze and Nam, H W and Pokharel, G and Wilson, Stephen D and Cho, Jun-Hyung and Moon, S J},
  journal={Physical Review B},
  volume={108},
  number={20},
  pages={205118},
  year={2023},
  publisher={APS}
}

@article{shi2025weak,
  title={Weak low-temperature ferromagnetism and linear magnetoresistance in {Lu$_{0.75}$Fe$_6$Sn$_6$} with a disordered {HfFe$_6$Ge$_6$}-type structure},
  author={Shi, Chenfei and Lin, Zhaodi and Liu, Qiyuan and Lyu, Junai and Xu, Xiaofan and Kang, Baojuan and Yang, Jin-Hu and Liu, Yi and Zhang, Jian and Cao, Shixun and Bao, Jin-Ke},
  journal={Journal of Magnetism and Magnetic Materials},
  volume={615},
  pages={172793},
  year={2025},
  publisher={Elsevier}
}

@article{konyk2020electrical,
    author = {Konyk, M and Romaka, L and Kuzhel, B and Stadnyk, Yu and Romaka, V V},
    title = {Electrical transport properties of {RCr$_6$Ge$_6$ (R=Y, Gd, Tb, Dy, Lu)} compounds},
    journal = {Visnyk of the Lviv University Chemistry Series},
    year = {2020},
    issue = {61},
    pages = {107-113},
    doi = {https://doi.org/10.30970/vch.6101.107}
}

@article{schobinger1997atomic,
  title={Atomic disorder and canted ferrimagnetism in the {TbCr$_6$Ge$_6$} compound. A neutron study},
  author={Schobinger-Papamantellos, P and Rodr{\'\i}guez-Carvajal, J and Buschow, K H J},
  journal={Journal of Alloys and Compounds},
  volume={255},
  number={1-2},
  pages={67--73},
  year={1997},
  publisher={Elsevier}
}

@article{schobinger1997ferrimagnetism,
  title={Ferrimagnetism and disorder in the {RCr$_6$Ge$_6$} compounds {(R=Dy, Ho, Er, Y): A} neutron study},
  author={Schobinger-Papamantellos, P and Rodr{\'\i}guez-Carvajal, J and Buschow, K H J},
  journal={Journal of Alloys and Compounds},
  volume={256},
  number={1-2},
  pages={92--96},
  year={1997},
  publisher={Elsevier}
}

@article{mulder1993Gd,
    author = {Mulder, F M and Thiel, R C and Brabers, J H V J and de Boer, F R and Buschow, K H J},
    title = {$^{155}${Gd M{\"o}ssbauer effect and magnetic properties of GdMn$_{6-x}$Cr$_x$Ge$_6$}},
    journal = {Journal of Alloys and Compounds},
    year = {1993},
    volume = {198},
    pages = {L1-L3},
    doi = {https://doi.org/10.1016/0925-8388(93)90130-F}
}

@article{brabers1994magnetic,
  title={Magnetic properties of {RCr$_6$Ge$_6$} compounds},
  author={Brabers, J H V J and Buschow, K H J and De Boer, F R},
  journal={Journal of Alloys and Compounds},
  volume={205},
  number={1-2},
  pages={77--80},
  year={1994},
  publisher={Elsevier}
}

@article{sato1983magnetic,
  title={Magnetic and electrical properties of {CrGe} and {Cr$_{11}$Ge$_8$}},
  author={Sato, Tetsuya and Sakata, Makoto},
  journal={Journal of the Physical Society of Japan},
  volume={52},
  number={5},
  pages={1807--1813},
  year={1983},
  publisher={The Physical Society of Japan}
}

@article{zagryazhskii1968magnetic,
  title={Magnetic susceptibility and electrical conductivity of the highest chromium germanide},
  author={Zagryazhskii, V L and Gel'd, P V and Shtol'ts, A K},
  journal={Soviet Physics Journal},
  volume={11},
  number={5},
  pages={23--25},
  year={1968},
  publisher={Springer}
}

@article{sato1988electrical,
  title={Electrical resistivity and magnetic susceptibility in the amorphous {Cr$_x$Ge$_{1-x}$} alloy system},
  author={Sato, T and Jono, A and Ohta, Eiji and Sakata, M},
  journal={Physical Review B},
  volume={38},
  number={16},
  pages={11741},
  year={1988},
  publisher={APS}
}

@article{moussa2019overview,
  title={Overview of the {U$_3$$T$Ge$_5$} family with {$T$= Ti, V, Cr, Mn, Zr, Nb, Mo, Hf, Ta and W: Nine} new members, phase formation, stability, structural and physical properties and electronic structures},
  author={Moussa, Chantal and Brisset, Nicolas and Chajewski, Grzegorz and Samsel-Czeka{\l}a, Ma{\l}gorzata and Boulet, Pascal and No{\"e}l, Henri and Pasturel, Mathieu and Pikul, Adam and Tougait, Olivier},
  journal={Journal of Solid State Chemistry},
  volume={277},
  pages={260--270},
  year={2019},
  publisher={Elsevier}
}

@article{kolenda1980esca,
  title={ESCA and magnetic studies of the {Cr-Ge} system},
  author={Kolenda, M and Stoch, J and Szytu{\l}a, A},
  journal={Journal of Magnetism and Magnetic Materials},
  volume={20},
  number={1},
  pages={99--106},
  year={1980},
  publisher={Elsevier}
}

@article{ghimire2012complex,
  title={Complex itinerant ferromagnetism in noncentrosymmetric {Cr$_{11}$Ge$_{19}$}},
  author={Ghimire, N J and McGuire, Michael A and Parker, David S and Sales, Brian C and Yan, J-Q and Keppens, V and Koehler, M and Latture, R M and Mandrus, D},
  journal={Physical Review B},
  volume={85},
  number={22},
  pages={224405},
  year={2012},
  publisher={APS}
}

@article{ji2025small,
  title={Small {Seebeck} in {A15}-type {Cr$_3$Ge} single crystals},
  author={Ji, Xiaoyu and Zhou, Xuebo and Zhu, Shilin and Ma, Fengcai and Li, Gang and Wu, Wei},
  journal={Europhysics Letters},
  year={2025}
}

@article{kresse1999ultrasoft,
  title={From ultrasoft pseudopotentials to the projector augmented-wave method},
  author={Kresse, Georg and Joubert, Daniel},
  journal={Physical Review B},
  volume={59},
  number={3},
  pages={1758},
  year={1999},
  publisher={APS}
}

@article{kresse1996efficient,
  title={Efficient iterative schemes for $ab initio$ total-energy calculations using a plane-wave basis set},
  author={Kresse, Georg and Furthm{\"u}ller, J{\"u}rgen},
  journal={Physical Review B},
  volume={54},
  number={16},
  pages={11169},
  year={1996},
  publisher={APS}
}

@article{kresse1993ab,
  title={$Ab initio$ molecular dynamics for open-shell transition metals},
  author={Kresse, Georg and Hafner, J},
  journal={Physical Review B},
  volume={48},
  number={17},
  pages={13115},
  year={1993},
  publisher={APS}
}

@article{perdew1996generalized,
  title={Generalized gradient approximation made simple},
  author={Perdew, John P and Burke, Kieron and Ernzerhof, Matthias},
  journal={Physical Review Letters},
  volume={77},
  number={18},
  pages={3865},
  year={1996},
  publisher={APS}
}

@article{venturini2006filling,
  title={Filling the {CoSn} host-cell: the {HfFe$_6$Ge$_6$}-type and the related structures},
  author={Venturini, Gerard},
  journal={Zeitschrift f{\"u}r Kristallographie-Crystalline Materials},
  volume={221},
  number={5-7},
  pages={511--520},
  year={2006},
  publisher={De Gruyter Oldenbourg}
}

@article{dudarev1998electron,
  title={Electron-energy-loss spectra and the structural stability of nickel oxide: {An LSDA+U} study},
  author={Dudarev, Sergei L and Botton, Gianluigi A and Savrasov, Sergey Y and Humphreys, C J and Sutton, Adrian P},
  journal={Physical Review B},
  volume={57},
  number={3},
  pages={1505},
  year={1998},
  publisher={APS}
}

@article{pokharel2021electronic,
  title={Electronic properties of the topological kagome metals {YV$_6$Sn$_6$} and {GdV$_6$Sn$_6$}},
  author={Pokharel, Ganesh and Teicher, Samuel M L and Ortiz, Brenden R and Sarte, Paul M and Wu, Guang and Peng, Shuting and He, Junfeng and Seshadri, Ram and Wilson, Stephen D},
  journal={Physical Review B},
  volume={104},
  number={23},
  pages={235139},
  year={2021},
  publisher={APS}
}

@article{xiao2024preparation,
  title={Preparation, crystal structure, and properties of the kagome metal {ThV$_6$Sn$_6$}},
  author={Xiao, Yusen and Chen, Yongliang and Ni, Hao and Li, Yong and Wen, Zhiwei and Cui, Yajing and Zhang, Yong and Liu, Shaohua and Wang, Cao and Zhong, Ruidan and others},
  journal={Inorganic Chemistry},
  volume={63},
  number={49},
  pages={23288--23295},
  year={2024},
  publisher={ACS Publications}
}

@article{pokharel2022highly,
  title={Highly anisotropic magnetism in the vanadium-based kagome metal {TbV$_6$Sn$_6$}},
  author={Pokharel, Ganesh and Ortiz, Brenden and Chamorro, Juan and Sarte, Paul and Kautzsch, Linus and Wu, Guang and Ruff, Jacob and Wilson, Stephen D},
  journal={Physical Review Materials},
  volume={6},
  number={10},
  pages={104202},
  year={2022},
  publisher={APS}
}

@article{clatterbuck1999magnetic,
  title={Magnetic properties of {RMn$_6$Sn$_6$ (R=Tb, Ho, Er, Tm, Lu)} single crystals},
  author={Clatterbuck, D M and Gschneidner Jr, K A},
  journal={Journal of Magnetism and Magnetic Materials},
  volume={207},
  number={1-3},
  pages={78--94},
  year={1999},
  publisher={Elsevier}
}

@article{kimura2006high,
  title={High-field magnetization of {RMn$_6$Sn$_6$} compounds with {R=Gd, Tb, Dy and Ho}},
  author={Kimura, S and Matsuo, A and Yoshii, S and Kindo, K and Zhang, L and Br{\"u}ck, E and Buschow, K H J and De Boer, F R and Lef{\`e}vre, C and Venturini, G},
  journal={Journal of Alloys and Compounds},
  volume={408},
  pages={169--172},
  year={2006},
  publisher={Elsevier}
}

@article{riberolles2024new,
  title={New insight into tuning magnetic phases of {$R$Mn$_6$Sn$_6$} kagome metals},
  author={Riberolles, Simon X M and Han, Tianxiong and Slade, Tyler J and Wilde, J M and Sapkota, A and Tian, Wei and Zhang, Qiang and Abernathy, D L and Sanjeewa, L D and Bud’ko, S L and Canfield, P C and McQueeney, R J and Ueland, B G},
  journal={npj Quantum Materials},
  volume={9},
  number={1},
  pages={42},
  year={2024},
  publisher={Nature Publishing Group UK London}
}

@article{ghimire2020competing,
  title={Competing magnetic phases and fluctuation-driven scalar spin chirality in the kagome metal {YMn$_6$Sn$_6$}},
  author={Ghimire, Nirmal J and Dally, Rebecca L and Poudel, L and Jones, D C and Michel, D and Magar, N Thapa and Bleuel, M and McGuire, Michael A and Jiang, J S and Mitchell, J F and Lynn, J W and Mazin, I I},
  journal={Science Advances},
  volume={6},
  number={51},
  pages={eabe2680},
  year={2020},
  publisher={American Association for the Advancement of Science}
}

@article{l2025high,
  title={High-field magnetic phase diagrams of the {$R$Mn$_6$Sn$_6$ ($R$= Gd--Tm)} kagome metals},
  author={L, Nil and Trevisan, Thais Victa and McQueeney, R J},
  journal={Physical Review B},
  volume={111},
  number={5},
  pages={054410},
  year={2025},
  publisher={APS}
}

@article{dhakal2021anisotropically,
  title={Anisotropically large anomalous and topological {Hall} effect in a kagome magnet},
  author={Dhakal, Gyanendra and Cheenicode Kabeer, Fairoja and Pathak, Arjun K and Kabir, Firoza and Poudel, Narayan and Filippone, Randall and Casey, Jacob and Pradhan Sakhya, Anup and Regmi, Sabin and Sims, Christopher and others},
  journal={Physical Review B},
  volume={104},
  number={16},
  pages={L161115},
  year={2021},
  publisher={APS}
}

@article{meier2023tiny,
  title={Tiny {Sc} allows the chains to rattle: {Impact} of {Lu} and {Y} doping on the charge-density wave in {ScV$_6$Sn$_6$}},
  author={Meier, William R and Madhogaria, Richa Pokharel and Mozaffari, Shirin and Marshall, Madalynn and Graf, David E and McGuire, Michael A and Arachchige, Hasitha W Suriya and Allen, Caleb L and Driver, Jeremy and Cao, Huibo and others},
  journal={Journal of the American Chemical Society},
  volume={145},
  number={38},
  pages={20943--20950},
  year={2023},
  publisher={ACS Publications}
}

@article{meier2025pressure,
  title={Pressure suppresses the density wave order in kagome metal {LuNb$_6$Sn$_6$}},
  author={Meier, William R and Graf, David E and Ortiz, Brenden R and Mozaffari, Shirin and Mandrus, David},
  journal={Physical Review Materials},
  volume={9},
  number={8},
  pages={L082001},
  year={2025},
  publisher={APS}
}

@article{arachchige2022charge,
  title={Charge density wave in kagome lattice intermetallic {ScV$_6$Sn$_6$}},
  author={Arachchige, Hasitha W Suriya and Meier, William R and Marshall, Madalynn and Matsuoka, Takahiro and Xue, Rui and McGuire, Michael A and Hermann, Raphael P and Cao, Huibo and Mandrus, David},
  journal={Physical Review Letters},
  volume={129},
  number={21},
  pages={216402},
  year={2022},
  publisher={APS}
}

@article{tuniz2023dynamics,
  title={Dynamics and resilience of the unconventional charge density wave in {ScV$_6$Sn$_6$} bilayer kagome metal},
  author={Tuniz, Manuel and Consiglio, Armando and Puntel, Denny and Bigi, Chiara and Enzner, Stefan and Pokharel, Ganesh and Orgiani, Pasquale and Bronsch, Wibke and Parmigiani, Fulvio and Polewczyk, Vincent and others},
  journal={Communications Materials},
  volume={4},
  number={1},
  pages={103},
  year={2023},
  publisher={Nature Publishing Group UK London}
}

@article{pokharel2023frustrated,
  title={Frustrated charge order and cooperative distortions in {ScV$_6$Sn$_6$}},
  author={Pokharel, Ganesh and Ortiz, Brenden R and Kautzsch, Linus and Gomez Alvarado, S J and Mallayya, Krishnanand and Wu, Guang and Kim, Eun-Ah and Ruff, Jacob P C and Sarker, Suchismita and Wilson, Stephen D},
  journal={Physical Review Materials},
  volume={7},
  number={10},
  pages={104201},
  year={2023},
  publisher={APS}
}

@article{hu2024phonon,
  title={Phonon promoted charge density wave in topological kagome metal {ScV$_6$Sn$_6$}},
  author={Hu, Yong and Ma, Junzhang and Li, Yinxiang and Jiang, Yuxiao and Gawryluk, Dariusz Jakub and Hu, Tianchen and Teyssier, J{\'e}r{\'e}mie and Multian, Volodymyr and Yin, Zhouyi and Xu, Shuxiang and others},
  journal={Nature Communications},
  volume={15},
  number={1},
  pages={1658},
  year={2024},
  publisher={Nature Publishing Group UK London}
}

@article{yi2024tuning,
  title={Tuning charge density wave of kagome metal {ScV$_6$Sn$_6$}},
  author={Yi, Changjiang and Feng, Xiaolong and Kumar, Nitesh and Felser, Claudia and Shekhar, Chandra},
  journal={New Journal of Physics},
  volume={26},
  number={5},
  pages={052001},
  year={2024},
  publisher={IOP Publishing}
}

@article{gonccalves1994ufe6ge6,
  title={{UFe$_6$Ge$_6$}: {A} new ternary magnetic compound},
  author={Gon{\c{c}}alves, A P and Waerenborgh, J C and Bonfait, G and Amaro, A and Godinho, M M and Almeida, M and Spirlet, J C},
  journal={Journal of Alloys and Compounds},
  volume={204},
  number={1-2},
  pages={59--64},
  year={1994},
  publisher={Elsevier}
}

@article{waerenborgh2005crystal,
  title={Crystal structure, $^{57}${Fe} {M{\"o}ssbauer} spectroscopy and magnetization of {U$_x$Fe$_6$Sn$_6$} (0${\leq}x{\leq}$0.6)},
  author={Waerenborgh, J C and Pereira, L C J and Gon{\c{c}}alves, A P and No{\"e}l, H},
  journal={Intermetallics},
  volume={13},
  number={5},
  pages={490--496},
  year={2005},
  publisher={Elsevier}
}

@article{buchholz1981intermetallische,
  title={Intermetallische Phasen mit {B35-{\"U}berstruktur} und Verwandtschaftsbeziehung zu {LiFe$_6$Ge$_6$}},
  author={Buchholz, Werner and Schuster, Hans-Uwe},
  journal={Zeitschrift f{\"u}r anorganische und allgemeine Chemie},
  volume={482},
  number={11},
  pages={40--48},
  year={1981},
  publisher={Wiley}
}

@misc{saint2025bruker,
  title={{\textsc{SAINT} V8.42}},
  author={{Bruker AXS SE}},
  year={2025},
}

@article{krause2015comparison,
  title={Comparison of silver and molybdenum microfocus {X-ray} sources for single-crystal structure determination},
  author={Krause, Lennard and Herbst-Irmer, Regine and Sheldrick, George M and Stalke, Dietmar},
  journal={Applied Crystallography},
  volume={48},
  number={1},
  pages={3--10},
  year={2015},
  publisher={International Union of Crystallography}
}

@article{sheldrick2015shelxt,
  title={{SHELXT}--Integrated space-group and crystal-structure determination},
  author={Sheldrick, George M},
  journal={Acta Crystallographica Section A},
  volume={71},
  number={1},
  pages={3--8},
  year={2015},
  publisher={International Union of Crystallography}
}

@article{sheldrick2015crystal,
  title={Crystal structure refinement with {SHELXL}},
  author={Sheldrick, George M},
  journal={Acta Crystallographica Section C},
  volume={71},
  number={1},
  pages={3--8},
  year={2015},
  publisher={International Union of Crystallography}
}

@article{groom2016cambridge,
  title={The {Cambridge} structural database},
  author={Groom, Colin R and Bruno, Ian J and Lightfoot, Matthew P and Ward, Suzanna C},
  journal={Structural Science},
  volume={72},
  number={2},
  pages={171--179},
  year={2016},
  publisher={International Union of Crystallography}
}

@article{guo2023triangular,
  title={Triangular {Kondo} lattice in {YbV$_6$Sn$_6$} and its quantum critical behavior in a magnetic field},
  author={Guo, Kaizhen and Ye, Junyao and Guan, Shuyue and Jia, Shuang},
  journal={Physical Review B},
  volume={107},
  number={20},
  pages={205151},
  year={2023},
  publisher={APS}
}

@article{meier2020flat,
  title={Flat bands in the {CoSn}-type compounds},
  author={Meier, William R and Du, Mao-Hua and Okamoto, Satoshi and Mohanta, Narayan and May, Andrew F and McGuire, Michael A and Bridges, Craig A and Samolyuk, German D and Sales, Brian C},
  journal={Physical Review B},
  volume={102},
  number={7},
  pages={075148},
  year={2020},
  publisher={APS}
}

@article{onuki1992magnetic,
  title={Magnetic and electrical properties of {U--Ge} intermetallic compounds},
  author={{\=O}nuki, Yoshichika and Ukon, Isamu and Won Yun, Sung and Umehara, Izuru and Satoh, Kazuhiko and Fukuhara, Tadashi and Sato, Hideyuki and Takayanagi, Shigeru and Shikama, Mikio and Ochiai, Akira},
  journal={Journal of the Physical Society of Japan},
  volume={61},
  number={1},
  pages={293--299},
  year={1992},
  publisher={The Physical Society of Japan}
}

@article{troc2002magnetotransport,
  title={Magnetotransport of compounds in the {U--Ge} system},
  author={Tro{\'c}, Robert and No{\"e}l, Henri and Boulet, Pascal},
  journal={Philosophical Magazine B},
  volume={82},
  number={7},
  pages={805--824},
  year={2002},
  publisher={Taylor \& Francis}
}

@article{venkatraman1985CrU,
    author = {Venkatraman, M and Neumann, J P and Peterson, D E},
    title = {The {Cr--U (Chromium--Uranium)} system},
    journal = {Bulletin of Alloy Phase Diagrams},
    year = {1985},
    volume = {6},
    pages = {425-429}
}

@article{malaman1997magnetic,
  title={Magnetic properties of {NdMn$_6$Sn$_6$} and {SmMn$_6$Sn$_6$} compounds from susceptibility measurements and neutron diffraction study},
  author={Malaman, B and Venturini, G and El Idrissi, B Chafik and Ressouche, E},
  journal={Journal of Alloys and Compounds},
  volume={252},
  number={1-2},
  pages={41--49},
  year={1997},
  publisher={Elsevier}
}

@article{venturini2001crystallographic,
  title={Crystallographic and magnetic properties of {TbFe$_6$Ge$_{6-x}$Ga$_x$} compounds (0.5${\le}x{\le}$3.5)},
  author={Venturini, G},
  journal={Journal of alloys and compounds},
  volume={329},
  number={1-2},
  pages={8--21},
  year={2001},
  publisher={Elsevier}
}

@article{ivantchev2000subgroupgraph,
  title={\textsc{SUBGROUPGRAPH}: a computer program for analysis of group--subgroup relations between space groups},
  author={Ivantchev, S and Kroumova, E and Madariaga, G and P{\'e}rez-Mato, J M and Aroyo, M I},
  journal={Applied Crystallography},
  volume={33},
  number={4},
  pages={1190--1191},
  year={2000},
  publisher={International Union of Crystallography}
}

@article{lou2025strongly,
  title={{Strongly Entangled Kondo and Kagome Lattices and the Emergent Magnetic Ground State in Heavy-Fermion Kagome Metal {YbV$_6$Sn$_6$}}},
  author={Lou, Rui and Mende, Max and Vocaturo, Riccardo and Zhang, Hao and Dong, Qingxin and Li, Man and Ding, Pengfei and Cheng, Erjian and Liao, Zhiguang and Zhang, Yu and Lin, Junfa and Firouzmandi, Reza and Kocsis, Vilmos and Corredor, Laura T and Prots, Yurii and Suvorov, Oleksandr and Jana, Anupam and Fujii, Jun and Vobornik, Ivana and Janson, Oleg and Zhu, Wenliang and van den Brink, Jeroen and Krellner, Cornelius and Pan, Minghu and Wang, Bosen and Xia, Tianlong and Cheng, Jinguang and Wang, Shancai and Felser, Claudia and B{\"u}chner, Bernd and Borisenko, Sergey and Yu, Rong and Vyalikh, Denis V and Fedorov, Alexander},
  journal={Physical Review Letters},
  volume={135},
  number={14},
  pages={146902},
  year={2025},
  publisher={APS}
}

@misc{lee2025coexisting,
  title={Coexisting Kagome and Heavy Fermion Flat Bands in {YbCr$_6$Ge$_6$}},
  author={Lee, Hanoh and Lyi, Churlhi and Lee, Taehee and Na, Hyeonhui and Kim, Jinyoung and Lee, Sangjae and Kim, Younsik and Rajapitamahuni, Anil and Kundu, Asish K and Vescovo, Elio and Park, Byeong-Gyu and Kim, Changyoung and Ahn, Charles H and Walker, Frederick J and Oh, {Ji Seop} and Jang, {Bo Gyu} and Kim, Youngkuk and Sohn, Byungmin and Park, Tuson},
  year={2025},
  eprint={2509.04902},
  archivePrefix={arXiv},
  primaryClass={cond-mat.str-el}
}

@misc{lv2026cooperative,
    title={Cooperative concurrence of $4f$ and $3d$ flat bands in kagome heavy-fermion metal {YbCr$_6$Ge$_6$}},
  author={Lv, Wenxin and Ma, Pengcheng and Wang, Tianqi and Tian, Shangjie and Ma, Ying and Wang, Shouguo and Zhang, Xiao and Liu, Zhonghao and Lei, Hechang},
  year={2026},
  eprint={2601.05829},
  archivePrefix={arXiv},
  primaryClass={cond-mat.str-el}
}

@article{wang2020experimental,
  title={Experimental observation of electronic structures of kagome metal {YCr$_6$Ge$_6$}},
  author={Wang, Pengdong and Wang, Yihao and Zhang, Bo and Li, Yuliang and Wang, Sheng and Wu, Yunbo and Zhu, Hongen and Liu, Yi and Zhang, Guobin and Liu, Dayong and Xiong, Yimin and Sun, Zhe},
  journal={Chinese Physics Letters},
  volume={37},
  number={8},
  pages={087102},
  year={2020},
  publisher={Chinese Physical Society and IOP Publishing Ltd}
}

@article{hoesch2017facility,
  title={A facility for the analysis of the electronic structures of solids and their surfaces by synchrotron radiation photoelectron spectroscopy},
  author={Hoesch, Moritz and Kim, T K and Dudin, P and Wang, H and Scott, S and Harris, P and Patel, S and Matthews, M and Hawkins, D and Alcock, S G and Richter, T and Mudd, J J and Basham, M and Pratt, L and Leicester, P and Longhi, E C and Tamai, A and Baumberger, F},
  journal={Review of Scientific Instruments},
  volume={88},
  number={1},
  year={2017},
  publisher={AIP Publishing}
}

\end{document}


\title{Supplemental Material}

\author{Z.~W.~Riedel}
\affiliation{Los Alamos National Laboratory, Los Alamos, New Mexico 87545, USA}

\author{{\color{black}P.~A.~E.~Murgatroyd}}
\affiliation{Los Alamos National Laboratory, Los Alamos, New Mexico 87545, USA}

\author{C.~S.~Kengle}
\affiliation{Los Alamos National Laboratory, Los Alamos, New Mexico 87545, USA}

\author{{\color{black}P.~M.~T.~Vianez}}
\affiliation{Los Alamos National Laboratory, Los Alamos, New Mexico 87545, USA}

\author{A.~Schmidt}
\affiliation{Bruker AXS, Madison, Wisconsin 53711, USA}

\author{{\color{black}X.~Du}}
\affiliation{{\color{black}Department of Applied Physics, Yale University, New Haven, Connecticut 06511, USA}}

\author{K.~Allen}
\affiliation{Los Alamos National Laboratory, Los Alamos, New Mexico 87545, USA}
\affiliation{
Department of Physics and
Astronomy, Rice University, Houston, Texas 77005, USA}

\author{{\color{black}T.~K.~Kim}}
\affiliation{{\color{black}Diamond Light Source Ltd., Harwell Science and Innovation Campus, Didcot, OX110DE, UK}}

\author{C.~Lane}
\affiliation{Los Alamos National Laboratory, Los Alamos, New Mexico 87545, USA}

\author{Ying Wai Li}
\affiliation{Los Alamos National Laboratory, Los Alamos, New Mexico 87545, USA}

\author{Jian-Xin Zhu}
\affiliation{Los Alamos National Laboratory, Los Alamos, New Mexico 87545, USA}

\author{J.~D.~Thompson}
\affiliation{Los Alamos National Laboratory, Los Alamos, New Mexico 87545, USA}

\author{F.~Ronning}
\affiliation{Los Alamos National Laboratory, Los Alamos, New Mexico 87545, USA}

\author{S.~M.~Thomas}
\affiliation{Los Alamos National Laboratory, Los Alamos, New Mexico 87545, USA}

\author{P.~F.~S.~Rosa}
\affiliation{Los Alamos National Laboratory, Los Alamos, New Mexico 87545, USA}

\author{E.~D.~Bauer}
\affiliation{Los Alamos National Laboratory, Los Alamos, New Mexico 87545, USA}

\maketitle

\section{Crystal Structure Refinement Details}
Additional refinement details for the room-temperature crystal structure of \UCG\ are provided in Table~\ref{tab:refinement}, and the corresponding atomic positions are in Table~\ref{tab:positions}.

\begin{table}[h]
    \centering
    \caption{Single-crystal XRD refinement details for \UCG}
    \begin{tabular}{l | c}
    \midrule
        Space Group~ & ~$C$2/$m$ \\
        $a$ (\AA)~ & ~5.1680(3) \\
        $b$ (\AA)~ & ~8.9509(5) \\
        $c$ (\AA)~ & ~4.1452(2) \\
        $\beta$ ($^\circ$)~ & ~90.023(2) \\
        $V$ (\AA$^3$)~ & ~191.750(18) \\
        $q$~ & ~[0.662(17)~\boldmath{$a^\ast$}, 0~\boldmath{$b^\ast$}, 0.501(8)~\boldmath{$c^\ast$}] \\
        Formula Unit, $Z$~ & ~\UCG, 1 \\
        $\mu$ (mm$^{-1}$)~ & ~52.183 \\
        2$\theta$ Range~ & ~[9.108, 60.846] \\
        Reflections~ & ~5733 \\
        GooF~ & ~1.207 \\
        $R_1$~ & ~1.40\% \\
        $wR_2$~ & ~3.78\% \\
        Largest peak/hole (e \AA$^{-3}$)~ & ~0.853/-0.891 \\
        \midrule
    \end{tabular}
    \label{tab:refinement}
\end{table}

\begin{table}[h]
    \centering
    \caption{Atomic positions for the \UCG\ average unit cell. Anisotropic displacement parameters are in units of \AA$^2$.}
    \begin{tabular}{l c c c c c c c c c c c c}
    \midrule Site & & x & y & z & occ. & U$_{11}$ & U$_{22}$ & U$_{33}$ & U$_{23}$ & U$_{13}$ & U$_{12}$ \\
    \midrule
        U1 & & 1/2 & 0 & 1/2 & 1/2 & 0.00126(17) & 0.00344(17) & 0.0025(2) & 0 & 0.00021(12) & 0 \\
        Cr1 & & 0 & 0 & 0 & 1 & 0.0019(3) & 0.0023(3) & 0.0027(4) & 0 & 0.0002(2) & 0 \\
        Cr2 & & 1/4 & 1/4 & 0 & 1 & 0.0007(2) & 0.0036(2) & 0.0026(3) & -0.00003(17) & 0.00019(17) & -0.00067(13) \\
        Ge1 & & 0.50009(11) & 0 & 0.1998(2) & 1/2 & 0.0021(3) & 0.0037(3) & 0.0033(3) & 0 & 0.0000(2) & 0 \\
        Ge2 & & 1/2 & 0.33334(3) & 1/2 & 1 & 0.00263(19) & 0.00467(19) & 0.0023(2) & 0 & 0.00006(14) & 0 \\
    \end{tabular}
    \label{tab:positions}
\end{table}

\clearpage
Additional refinement details for the room-temperature crystal structure of \LCG\ are provided in Table~\ref{tab:refinement_Lu}, and the corresponding atomic positions are in Table~\ref{tab:positions_Lu}. The largest residual electron density (``largest peak") was present at the unit cell edge, similar to that reported for Sn disorder in the common SmMn$_6$Sn$_6$ structure type \cite{malaman1997magnetic}. However, the largest peak was small for a heavy-element system (2.4~$e$/{\AA}$^3$), and no residual density was present at (0,0,0.5), where one would expect corresponding disorder of the Lu site. A refinement with Ge split between the primary site and the largest peak site converged to nonphysical occupancies ($>100$\% for the primary site) and, thus, did not improve on a fully ordered $P6/mmm$ structural model.

\begin{table}[ht]
    \centering
    \caption{Single-crystal XRD refinement details for \LCG}
    \begin{tabular}{l | c}
    \midrule
        Space Group~ & ~$P6/mmm$ \\
        $a=b$ (\AA)~ & ~5.1509(3) \\
        $c$ (\AA)~ & ~8.2654(8) \\
        $V$ (\AA$^3$)~ & ~189.92(3) \\
        Formula Unit, $Z$~ & ~\LCG, 1 \\
        $\mu$ (mm$^{-1}$)~ & ~44.314 \\
        2$\theta$ Range~ & ~[4.92, 82.22] \\
        Reflections~ & ~9912 \\
        GooF~ & ~1.302 \\
        $R_1$~ & ~2.29\% \\
        $wR_2$~ & ~5.84\% \\
        Largest peak/hole (e \AA$^{-3}$)~ & ~2.327/-1.568 \\
        \midrule
    \end{tabular}
    \label{tab:refinement_Lu}
\end{table}

\begin{table}[ht]
    \centering
    \caption{Atomic positions for the \LCG\ unit cell. Anisotropic displacement parameters are in units of \AA$^2$.}
    \begin{tabular}{l c c c c c c c c c c c c}
    \midrule Site & & x & y & z & occ. & U$_{11}$ & U$_{22}$ & U$_{33}$ & U$_{23}$ & U$_{13}$ & U$_{12}$ \\
    \midrule
        Lu1 & & 0 & 0 & 0 & 1 & 0.0011(3) & 0.0011(3) & 0.0094(4) & 0 & 0 & 0.00053(14) \\
        Cr1 & & 1/2 & 1/2 & 0.75035(8) & 1 & 0.0016(4) & 0.0016(4) & 0.0075(5) & 0 & 0 & 0.0012(3) \\
        Ge1 & & 0 & 0 & 0.65416(13) & 1 & 0.0008(4) & 0.0008(4) & 0.0080(5) & 0 & 0 & 0.00039(18) \\
        Ge2 & & 2/3 & 1/3 & 0 & 1 & 0.0018(3) & 0.0018(3) & 0.0069(5) & 0 & 0 & 0.00088(17) \\
        Ge3 & & 1/3 & 2/3 & 0.5 & 1 & 0.0036(4) & 0.0036(4) & 0.0063(5) & 0 & 0 & 0.00182(18) \\
    \end{tabular}
    \label{tab:positions_Lu}
\end{table}

\clearpage
\section{Unit cell comparisons}
\begin{figure}[h]
    \centering
    \includegraphics[width=0.6\columnwidth]{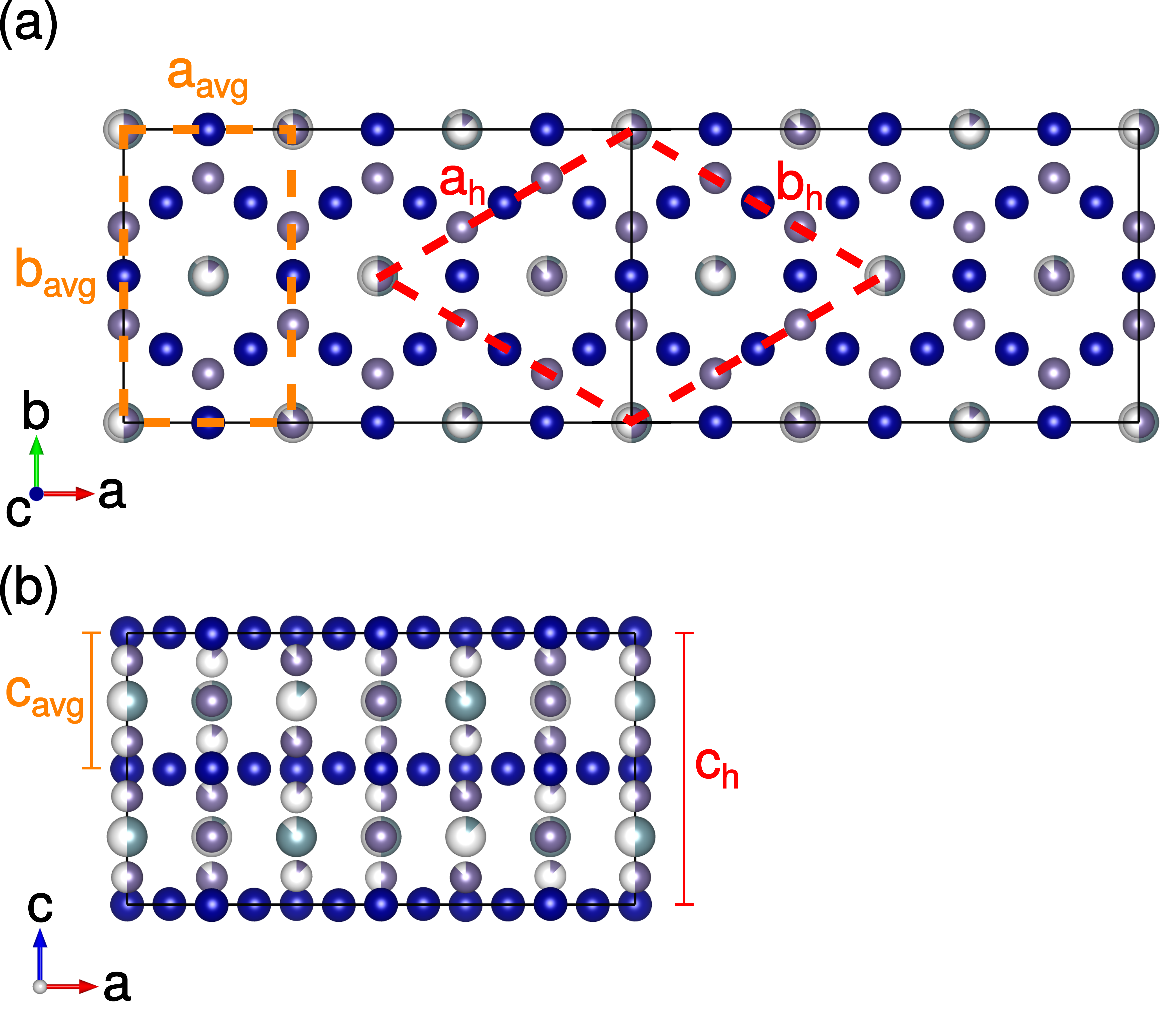}
    \caption{The relationships between the average monoclinic cell (``avg" subscript), the approximate hexagonal supercell (``h" subscript), and the approximate monoclinic supercell (``sup" subscript) are depicted. The axes refer to the monoclinic supercell. The colors follow the main text (U: teal, Cr: blue, Ge: purple). (a) Two monoclinic supercells viewed down the $c$ axis. (b) One monoclinic supercell viewed down the $b$ axis.}
    \label{fig:cell_relations}
\end{figure}

In the main text, multiple unit cells for \UCG\ are discussed. The first is the average cell of the modulated crystal structure with $C2/m$ space-group symmetry. This is a well-supported average cell with refinement details provided in Tables~\ref{tab:refinement} and \ref{tab:positions}. If instead the modulation and average-cell diffraction reflections are both considered during the refinement, the reflections suggest $P6_3/mmc$ symmetry. The resulting $R_1$ is 2.6\%, and the resulting $wR_2$ is 8.7\%. Both are worse than refinements to only the average-cell reflections because the refinement is not appropriately accounting for the modulated reflections' reduced intensity. Still, the fit may approximate the real-space modulation. $P6_3/mmc$ is a supergroup of $C2/m$ \cite{ivantchev2000subgroupgraph}, so the unit cell can be transformed from the hexagonal supercell lattice [(\textbf{a$_\mathrm{h}$},\textbf{b$_\mathrm{h}$},\textbf{c$_\mathrm{h}$})] to the monoclinic supercell lattice  presented in the main text [(\textbf{a$_\mathrm{sup}$},\textbf{b$_\mathrm{sup}$},\textbf{c$_\mathrm{sup}$})] by Eq.~\ref{eq:transform} without breaking any crystal symmetry. The resulting supercell corresponds to a 3$\times$1$\times$2 supercell of the refined average cell, as expected for $\textbf{q}$=(2/3,0,1/2). The periodicity of the supercell's alternating polyhedra matches the modulation wavevector, and the resulting 90$^\circ$ $\beta$ angle is reasonable based on the average-cell refinement's error bars. Further refining the data in the monoclinic supercell did not provide significant improvement, so the refined site occupations from the hexagonal cell fit were retained. The relationship between the lattice parameter magnitudes of the monoclinic average cell, the monoclinic supercell, and the hexagonal supercell are provided by Eq.~\ref{eq:cell_mag}. Figure~\ref{fig:cell_relations} shows the spatial relationships between the three cells.

\begin{equation}
    \label{eq:transform}
    (\textbf{a$_\mathrm{sup}$},\textbf{b$_\mathrm{sup}$},\textbf{c$_\mathrm{sup}$}) = (\textbf{a$_\mathrm{h}$},\textbf{b$_\mathrm{h}$},\textbf{c$_\mathrm{h}$})
    \begin{pmatrix}
    -1 & -1 & 0\\
    1 & -1 & 0\\
    0 & 0 & 1
    \end{pmatrix}
\end{equation}

\begin{equation}
\label{eq:cell_mag}
\begin{split}
    &|\textbf{a$_\mathrm{sup}$}|=3~|\textbf{a$_\mathrm{avg}$}|=\sqrt{3}~|\textbf{a$_\mathrm{h}$}| \\
    &|\textbf{b$_\mathrm{sup}$}|=|\textbf{b$_\mathrm{avg}$}|=|\textbf{b$_\mathrm{h}$}| \\
    &|\textbf{c$_\mathrm{sup}$}|=2~|\textbf{c$_\mathrm{avg}$}|=|\textbf{c$_\mathrm{h}$}|
\end{split}
\end{equation}

\clearpage



\clearpage
{\color{black}
\section{Magnetic Data Comparison for Multiple Crystals}
Magnetic property data is provided for two crystals to compare with the main text data (Fig.~\ref{fig:compare}). Together, the significant changes in the magnetic susceptibility [$\chi$($T$)] baseline above 30 K, the magnetic susceptibility jump below 30 K, and the curvature of the magnetization [$M$($H$)] below $\sim$0.2~T in C3 suggest that the transition is not intrinsic. The insets of the magnetization data show the more pronounced curvature and finite remnant magnetization of C3, which also has a larger high-temperature susceptibility baseline.
Paired with the ARPES data that suggests itinerant uranium $f$ electron weight, we conclude that the 30~K transition is due to magnetic impurities, and \UCG\ is paramagnetic.

\begin{figure}[h]
    \centering
    \subfloat{\includegraphics[width=0.4\columnwidth]{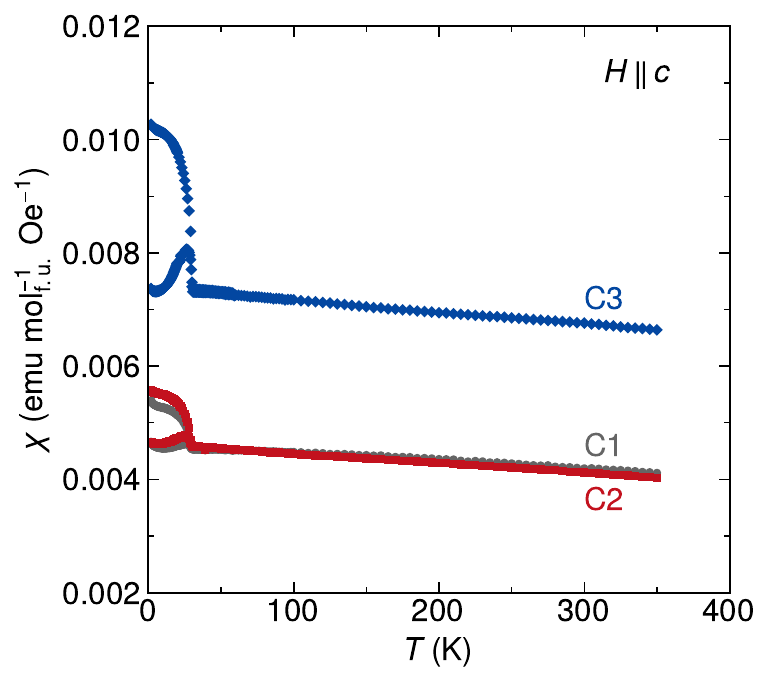}}
    \hspace{1.5em}
    \subfloat{\includegraphics[width=0.4\columnwidth]{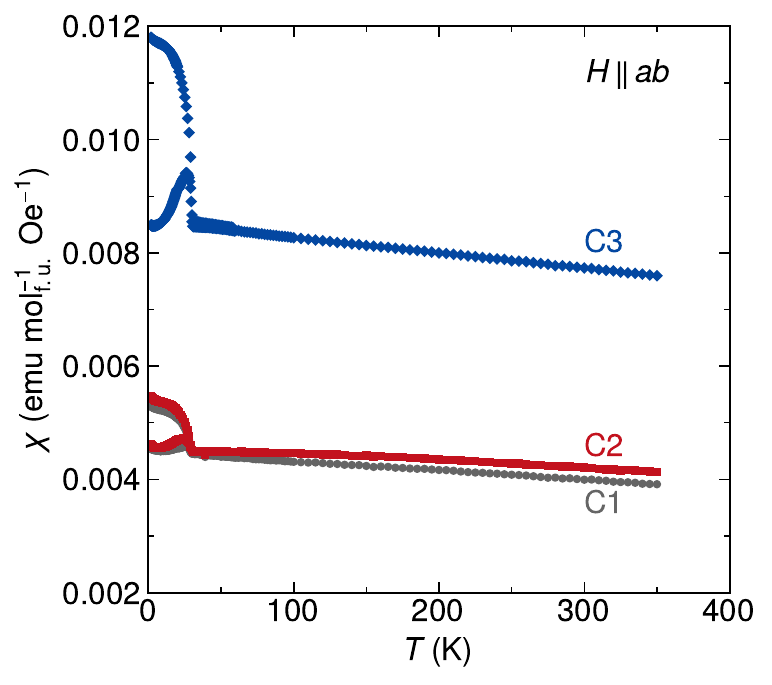}}
    \\   
    \subfloat{\includegraphics[width=0.4\columnwidth]{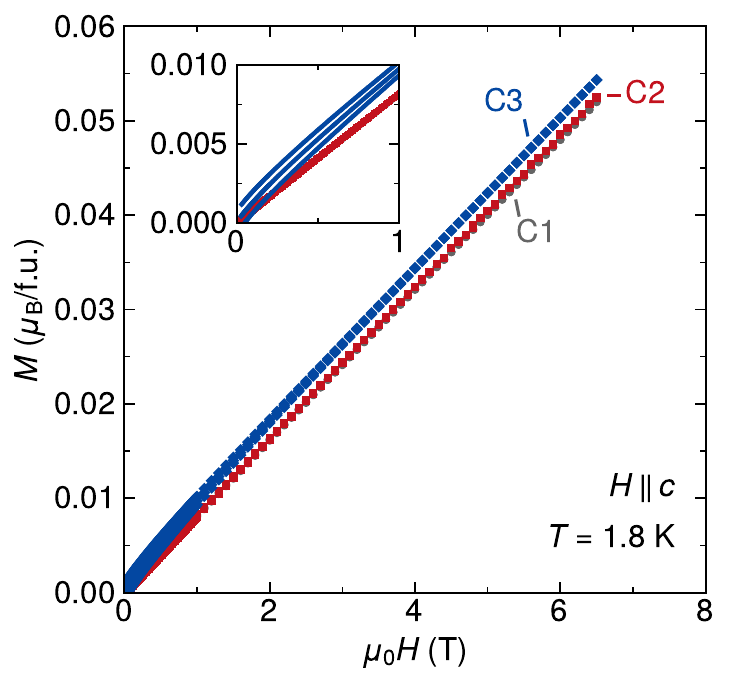}}  
    \hspace{1.5em}
    \subfloat{\includegraphics[width=0.4\columnwidth]{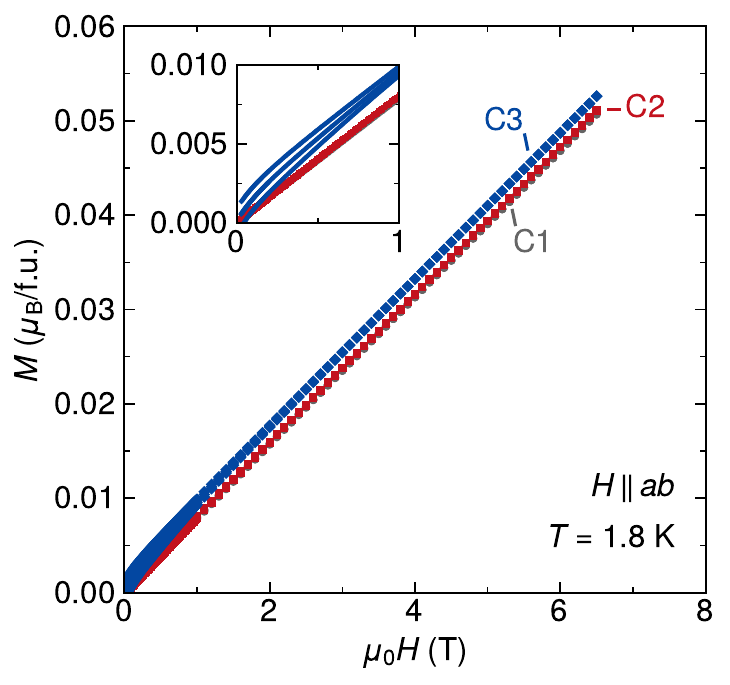}}
    \caption{Anisotropic magnetic susceptibility and magnetization data are presented for three \UCG\ crystals. Data for crystal 1 (``C1", gray circles) are also in the main text. Crystal 2 (``C2", red squares) is from the same batch as C1, and crystal 3 (``C3", blue diamonds) is from a separate batch.}
    \label{fig:compare}
\end{figure}
}

\clearpage
\section{Impurity Transitions in the Heat Capacity}
\begin{figure}[h]
    \centering
    \includegraphics[width=0.45\columnwidth]{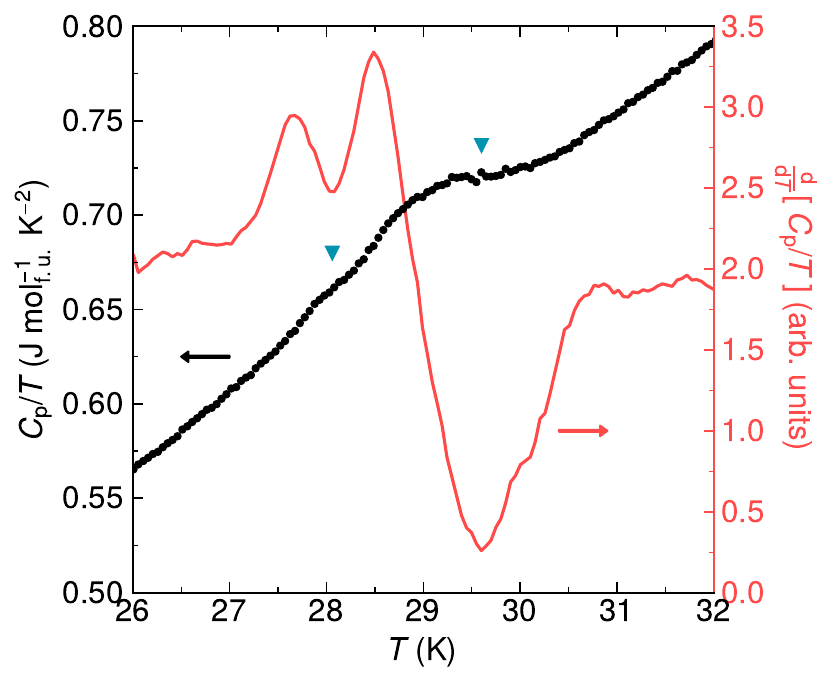}
    \caption{A subset of the $C_\mathrm{p}/T$ data for \UCG\ is shown along with its first derivative with respect to temperature. Two small features are marked with triangles.}
    \label{fig:hc_pts}
\end{figure}

The heat capacity of \UCG\ contains two features at 28.1 and 29.6~K, as defined by valleys in the first temperature derivative (Fig.~\ref{fig:hc_pts}). We attribute the features to the impurity phase discussed in the main text.

\section{L\lowercase{u}C\lowercase{r}$_6$G\lowercase{e}$_6$ Heat Capacity Comparison}
\begin{figure*}[h]
    \centering
    \subfloat{\includegraphics[width=0.45\columnwidth]{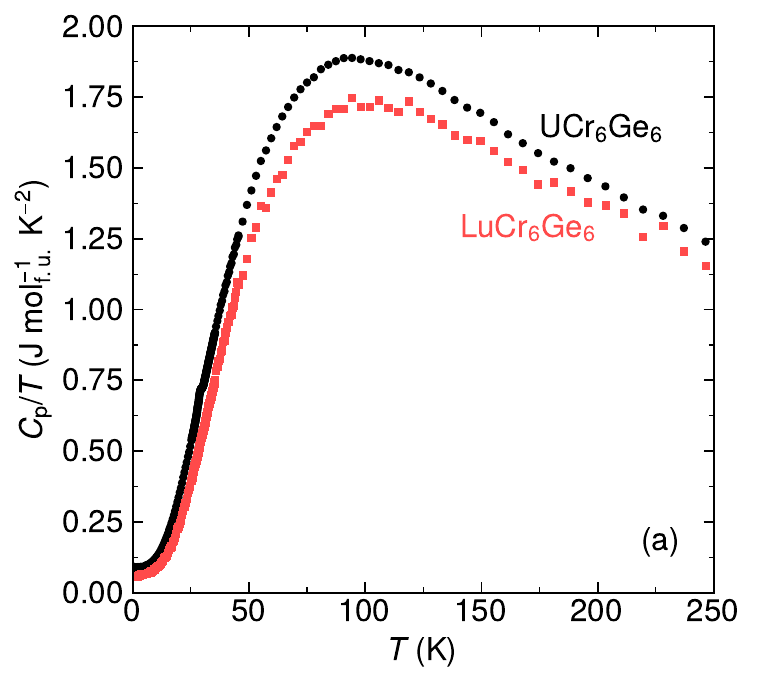}}
    \hspace{1em}
    \subfloat{\includegraphics[width=0.438\columnwidth]{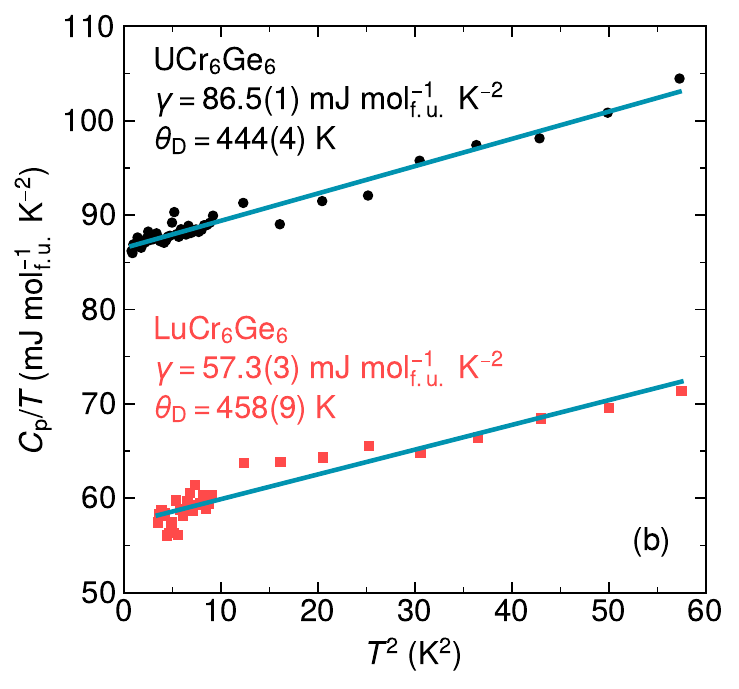}}
    \caption{(a) The zero-field heat capacities divided by temperature for \UCG\ and \LCG\ are shown along with (b) their fit Sommerfeld coefficients and Debye temperatures.}
    \label{fig:Lu_Cp}
\end{figure*}

Figure~\ref{fig:Lu_Cp}(a) compares the heat capacities of \UCG\ and \LCG, and Figure~\ref{fig:Lu_Cp}(b) compares the Sommerfeld coefficient and Debye temperature derived for both from low-temperature data (see main text for fit details). The \LCG\ electronic heat capacity stems primarily from chromium state contributions at the Fermi level, and the larger \UCG\ value suggests additional uranium 5$f$ state contributions.

\clearpage
\section{Additional DFT calculations}
\begin{figure*}[h]
    \centering
    \subfloat{\includegraphics[width=0.45\columnwidth]{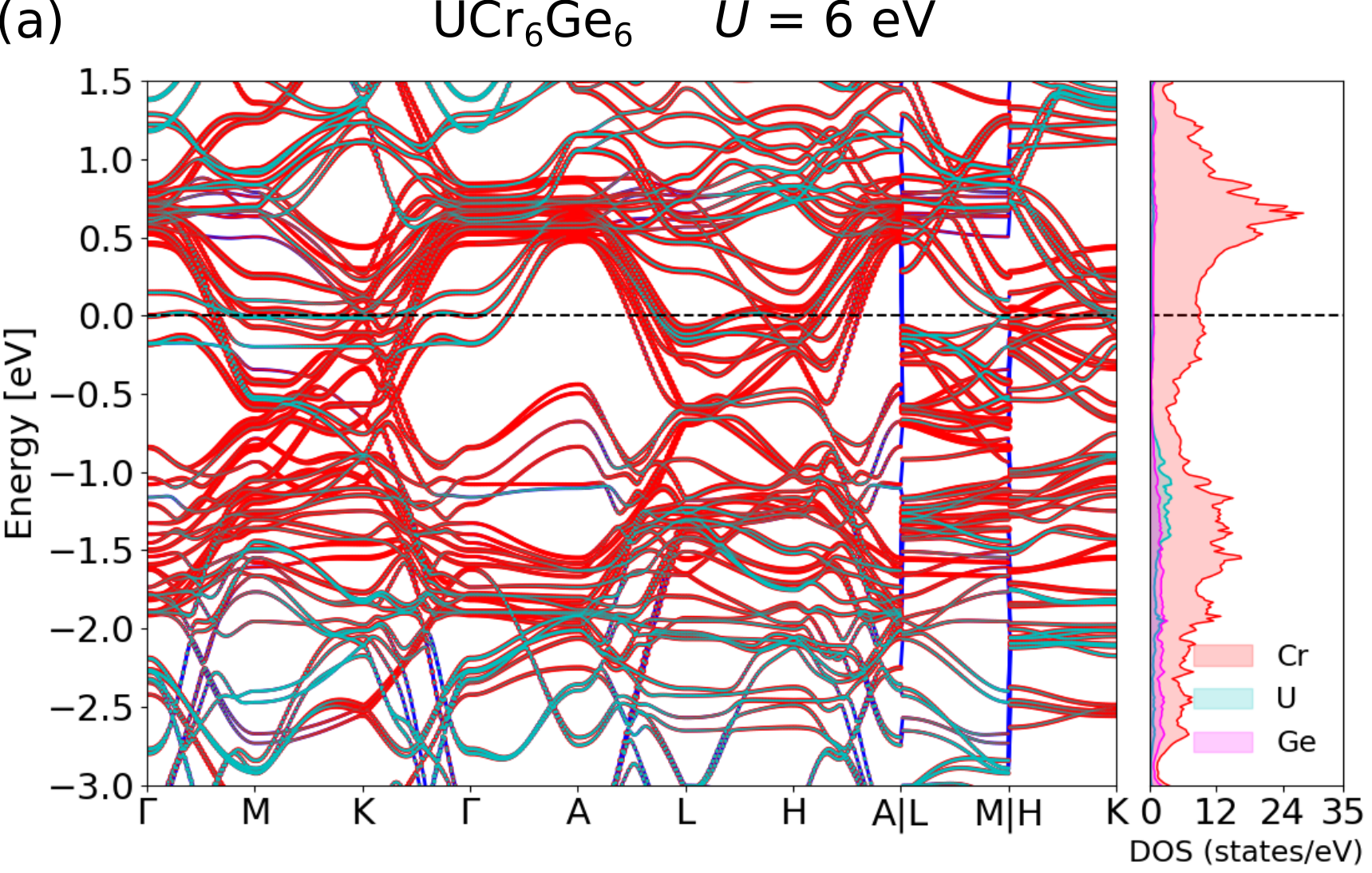}}
    \hspace{1em}
    \subfloat{\includegraphics[width=0.45\columnwidth]{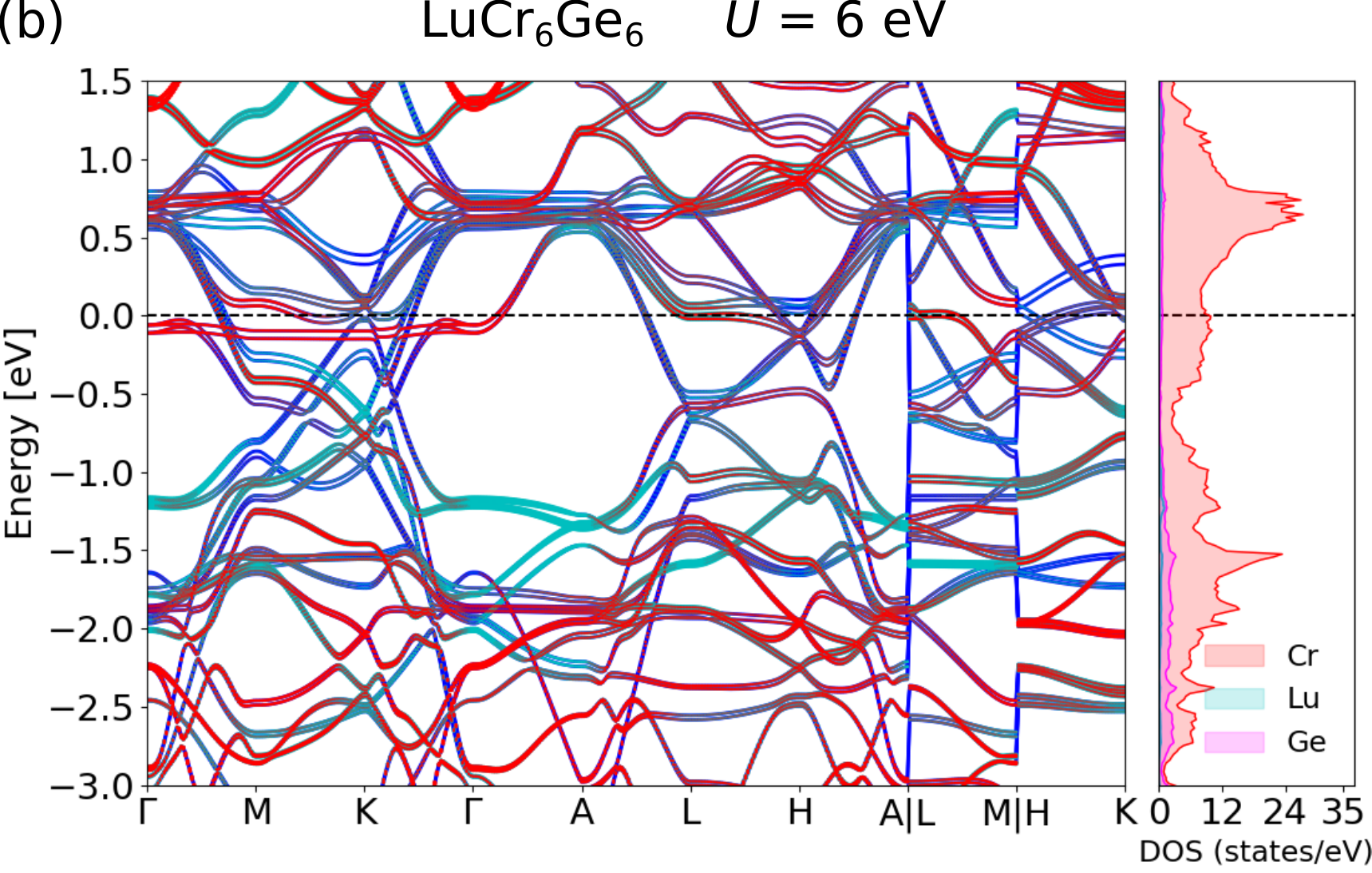}}
    \caption{Calculated electronic band structures of (a) \UCG\ and (b) \LCG\ with an effective Hubbard $U$}
    \label{fig:DFT_U}
\end{figure*}

DFT calculations of the electronic band structure of \UCG\ [Fig.~\ref{fig:DFT_U}(a)] and \LCG\ [Fig.~\ref{fig:DFT_U}(b)] are provided for an effective Hubbard $U$ of 6~eV. The $U$ localizes the uranium 5$f$ states, which may be itinerant in the real material.

{\color{black}\section{Additional ARPES Data}

Due to the three-dimensional electronic structure of \UCG, photon energy dependent measurements were performed to determine the relevant $k_\mathrm{z}$ planes. The sample was aligned such that the $\Gamma(A)–M(L)$ high-symmetry direction was parallel to the analyzer slit. Measurements were carried out over the photon-energy range h$\nu = 50–120$~eV using linear horizontal polarization. The data were converted to $k_\mathrm{z}$ using a nearly-free-electron final-state model with an inner potential $V_\mathrm{0}=18$~eV, determined iteratively. The resulting $k_\mathrm{x}-k_\mathrm{z}$ constant-energy surface at $E-E_\mathrm{F} = 0.120 \pm 0.025$~eV is shown in Fig.~\ref{fig:arpes_xtra}. The photon energies highlighted in the main text (92 and 108~eV) lie near the $k_\mathrm{z}=\pi$ and $k_\mathrm{z}=0$ planes, respectively. These energies were chosen to leverage matrix-element effects: 92~eV is near the uranium 5$f$ Cooper minimum, where 5$f$ photoemission intensity is suppressed, while 108~eV corresponds to the uranium O-edge ($5d \rightarrow 5f$) resonance, where 5$f$ spectral weight is enhanced.

\begin{figure}[h]
    \centering
    \includegraphics[width=0.6\columnwidth]{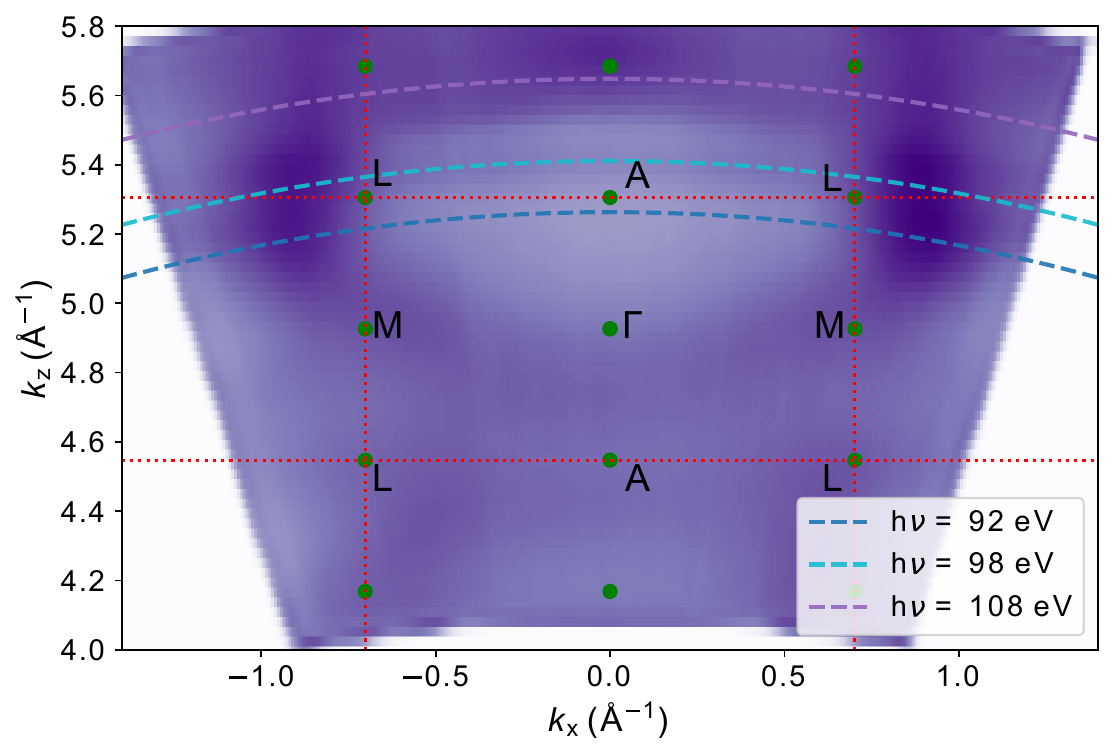}
    \caption{$k_\mathrm{x}-k_\mathrm{z}$ constant-energy surface at $E-E_\mathrm{F} =0.120 \pm 0.025$~eV obtained from photon energy dependent ARPES measurements (h$\nu =50–120$~eV, linear horizontal polarization, $T=8$~K) along the $\Gamma(A)–M(L)$ high-symmetry direction. Red dashed lines indicate Brillouin zone boundaries. High-symmetry points are marked in green. Blue, cyan, and purple dashed curves denote the photon energies used in the main text ($h\nu=92$, 98, and 108~eV, respectively).}
    \label{fig:arpes_xtra}
\end{figure}
}

\section{Additional Resistivity Data}

\begin{figure}[h]
    \centering
    \includegraphics[width=0.435\columnwidth]{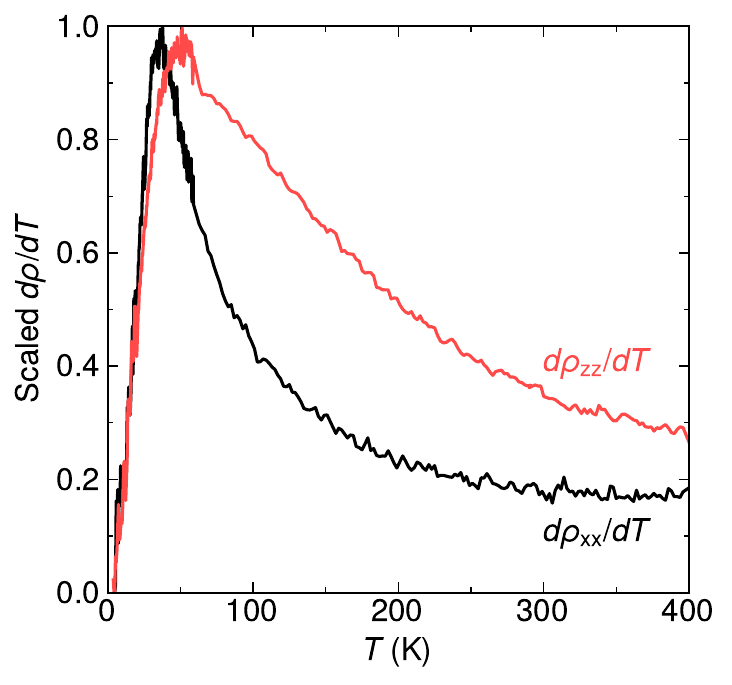}
    \caption{The first derivative of the longitudinal resistivity is plotted for an electrical current applied within the kagome plane ($d\rho_\mathrm{xx}/dT$) and perpendicular to it ($d\rho_\mathrm{zz}/dT$). The derivatives are scaled to the range [0,1] for comparison.}
    \label{fig:dRdT}
\end{figure}

Figure~\ref{fig:dRdT} shows the first derivative of the longitudinal resistivity with respect to temperature for \UCG. Neither orientation's data includes a peak that indicates a magnetic transition. Rather, they both show a hump at an inflection point in the resistivity.

\clearpage
{\color{black}Additional magnetoresistance and Hall resistivity data for \UCG\ are included in Figures~\ref{fig:xtra_MR} and \ref{fig:hall}. The crystals used were the same as those used for the main text figures. Figure~\ref{fig:xtra_MR} compares the 2~K magnetoresistance in the main text (current along the $c$ axis and the applied magnetic field within the kagome plane) to data collected at higher temperatures. The initially negative magnetoresistance approaches zero with increasing temperature. Figure~\ref{fig:hall}(a) shows the Hall resistivity ($\rho_\mathrm{yx}$) with a magnetic field applied perpendicular to the kagome plane for the same temperatures. Data were collected at each temperature after the same zero-field cooling temperature profile with positive and negative magnetic field data antisymmetrized to eliminate contributions from voltage lead mismatch [($R_{H>0}-R_{H<0}$)/2]. The slightly negative slope indicates dominant electron carriers in a one-band approximation. Figure~\ref{fig:hall}(b) shows the corresponding carrier concentration derived from the relationship $n_e=1/(R_Hq)$, where $R_H$ is the Hall coefficient [linear slope of $\rho_\mathrm{yx}(H)$] and $q$ is the charge of an electron.

\begin{figure}[h]
    \centering
    \includegraphics[width=0.435\columnwidth]{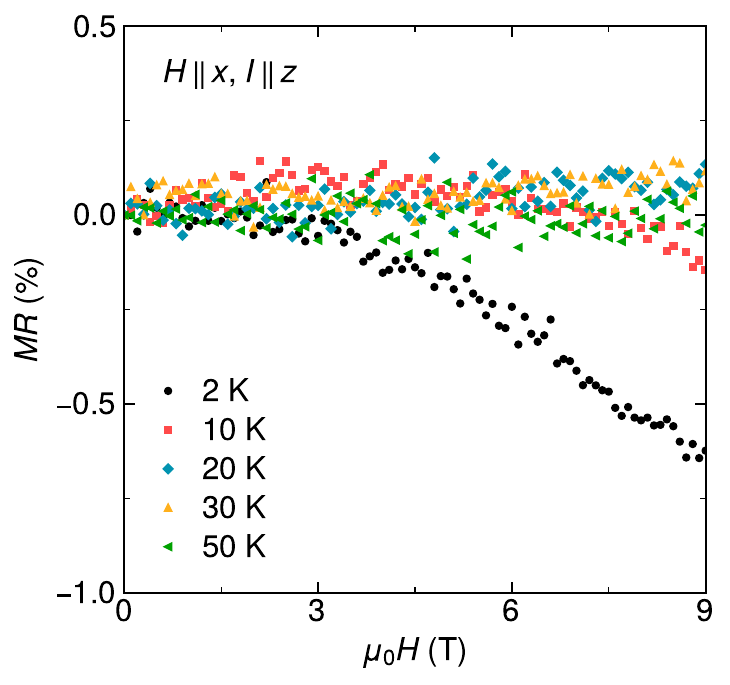}
    \caption{Magnetoresistance at 2, 10, 20, 30, and 50~K for \UCG, where $x$ and $y$ are perpendicular directions within the kagome plane and $z$ is along the crystallographic $c$ axis}
    \label{fig:xtra_MR}
\end{figure}

\begin{figure*}[h]
    \centering
    \subfloat{\includegraphics[width=0.45\columnwidth]{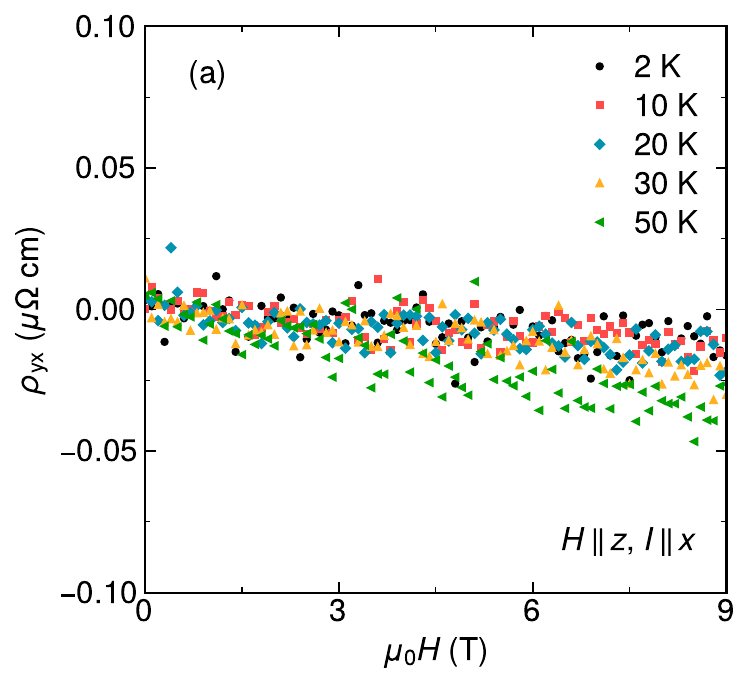}}
    \hspace{1em}
    \subfloat{\includegraphics[width=0.435\columnwidth]{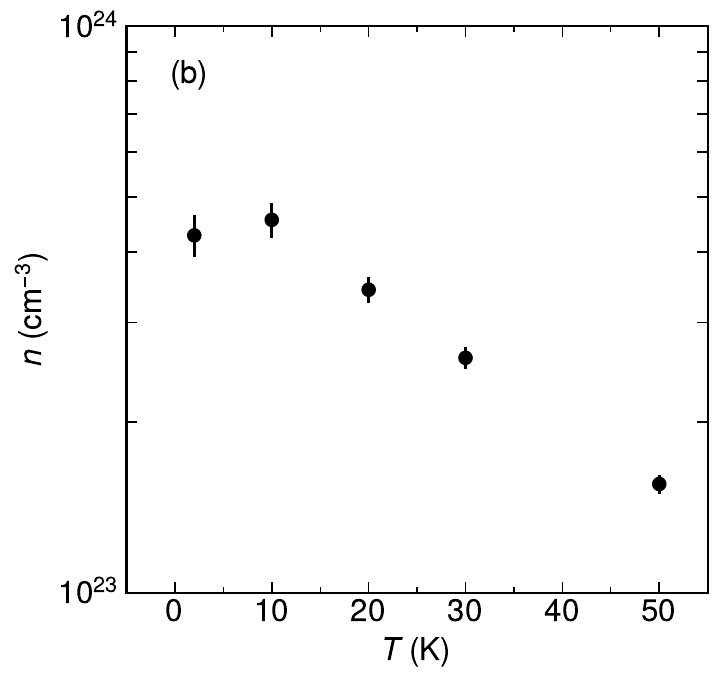}}
    \caption{(a) Hall resistivity at 2, 10, 20, 30, and 50~K for \UCG, where $x$ and $y$ are perpendicular directions within the kagome plane and $z$ is along the crystallographic $c$ axis (b) Corresponding electron carrier concentration in the one-band approximation with error bars representing the standard error associated with the least-squares fit}
    \label{fig:hall}
\end{figure*}
}

\clearpage
\bibliography{UCr6Ge6.bib}